\documentclass[sigconf, nonacm]{acmart}

\newcommand\vldbdoi{XX.XX/XXX.XX}
\newcommand\vldbpages{XXX-XXX}
\newcommand\vldbvolume{16}
\newcommand\vldbissue{1}
\newcommand\vldbyear{2023}
\newcommand\vldbauthors{\authors}
\newcommand\vldbtitle{\shorttitle} 
\newcommand\vldbavailabilityurl{https://github.com/ISG-ICS/QueryBooster}
\newcommand\vldbpagestyle{plain} 

\usepackage{balance}

\usepackage{soul}
\usepackage{listings}
\usepackage{amsmath,nicefrac,pifont}
\DeclareMathOperator*{\argmax}{arg\,max}
\usepackage{mathtools}

\usepackage[linesnumbered,ruled]{algorithm2e}
\usepackage{subcaption}
\usepackage{wrapfig}
\usepackage{multirow}

\definecolor{Yellow}{rgb}{255, 255, 0}
\definecolor{Blue}{rgb}{0, 0, 255}
\definecolor{Red}{rgb}{255, 0, 0}
\definecolor{OliveGreen}{HTML}{3C8031}
\definecolor{Indigo}{HTML}{6F00FF}
\definecolor{Brown}{rgb}{0.59, 0.29, 0.0}
\definecolor{CornellRed}{rgb}{0.7, 0.11, 0.11}

\newcommand{\rmeta}[1]{#1}
\newcommand{\rone}[1]{#1}
\newcommand{\rtwo}[1]{#1}
\newcommand{\rfour}[1]{#1}
\newcommand{\boldstart}[1]{\textbf{#1}}
\newcommand{\sysname}{{\sf QueryBooster}\xspace}
\newcommand{\langname}{{\sf VarSQL}\xspace}
\lstdefinelanguage{SQL}{%
  alsoletter={-_:\$\\},
  morekeywords={
    select,
    from,
    where,
    and,
    or,
    in,
    as,
    group,
    by,
    join,
    on,
    with,
    update,
    set,
    delete,
    insert,
    limit,
    having,
    is,
    distinct,
    is_pfkey,
    isa,
    not,
    strpos,
    lower,
    ilike,
    Rule,
    Pattern,
    Constraints,
    Replacement,
    Actions,
    adddate,
    date_format,
    interval,
    second,
    timestamp
  },
  emph={},  % highlighting
  emphstyle=\slshape\bfseries\color{blue},
  moredelim=**[is][\bfseries\color{red}]{@}{@}, % highlighting
  moredelim=**[is][\bfseries\color{blue}]{&}{&}, % highlighting
  basicstyle=\sf,
  keywordstyle=\textbf,
  identifierstyle=\texttt,
  morecomment=[s]{/*}{*/},
  commentstyle=\slshape\bfseries\color{blue},
  sensitive=false,
  literate={<=}{{\litleq}}1 {>=}{{\litgeq}}1 {//}{{\litdoubleslash}}1 %%{::}{{\litdoublecolon}}1%
}[keywords,comments,strings]

\lstnewenvironment{sql}[1]
   {\lstset{language=SQL,columns=flexible,basicstyle=\footnotesize,basewidth={.3em,.2em},mathescape=true,title=#1}}
   {}
\lstnewenvironment{sqlnotitle}
   {\lstset{language=SQL,columns=flexible,basicstyle=\linespread{1.24}\footnotesize,basewidth={.3em,.2em},mathescape=true}}
   {}
\lstnewenvironment{rule-example}
   {\lstset{language=SQL,columns=flexible,basicstyle=\linespread{1.13}\small,basewidth={.3em,.2em},mathescape=true}}
   {}
\SetCommentSty{mycommfont}
\SetKwInOut{Parameter}{Parameters}

\usepackage{tikz}
\newcommand*\circled[1]{\tikz[baseline=(char.base)]{
            \node[shape=circle,draw,inner sep=1pt] (char) {#1};}}

\begin{document}
\title{\sysname: Improving SQL Performance Using Middleware Services for Human-Centered Query Rewriting}

%%
%% The "author" command and its associated commands are used to define the authors and their affiliations.
\author{Qiushi Bai, Sadeem Alsudais, Chen Li}
\affiliation{%
  \institution{Department of Computer Science, UC Irvine, CA 92697, USA}
}
\email{{qbai1,salsudai}@uci.edu, chenli@ics.uci.edu}

\begin{abstract}
SQL query performance is critical in database applications, and query rewriting is a technique that transforms an original query into an equivalent query with a better performance.  In a wide range of database-supported systems, there is a unique problem where both the application and database layer are black boxes, and the developers need to use their knowledge about the data and domain to rewrite queries sent from the application to the database for better performance.  Unfortunately, existing solutions do not give the users enough freedom to express their rewriting needs.  To address this problem, we propose \sysname, a novel middleware-based service architecture for human-centered query rewriting, where users can use its expressive and easy-to-use rule language (called \langname) to formulate rewriting rules based on their needs.  It also allows users to express rewriting intentions by providing examples of the original query and its rewritten query. \sysname automatically generalizes them to rewriting rules and suggests high-quality ones.  We conduct a user study to show the benefits of \langname to formulate rewriting rules.  Our experiments on real and synthetic workloads show the effectiveness of the rule-suggesting framework and the significant advantages of using \sysname for human-centered query rewriting to improve the end-to-end query performance.
\end{abstract}

\maketitle

\pagestyle{\vldbpagestyle}
\begingroup\small\noindent\raggedright\textbf{PVLDB Reference Format:}\\
\vldbauthors. \vldbtitle. PVLDB, \vldbvolume(\vldbissue): \vldbpages, \vldbyear.\\
\href{https://doi.org/\vldbdoi}{doi:\vldbdoi}
\endgroup
\begingroup
\renewcommand\thefootnote{}\footnote{\noindent
This work is licensed under the Creative Commons BY-NC-ND 4.0 International License. Visit \url{https://creativecommons.org/licenses/by-nc-nd/4.0/} to view a copy of this license. For any use beyond those covered by this license, obtain permission by emailing \href{mailto:info@vldb.org}{info@vldb.org}. Copyright is held by the owner/author(s). Publication rights licensed to the VLDB Endowment. \\
\raggedright Proceedings of the VLDB Endowment, Vol. \vldbvolume, No. \vldbissue\ %
ISSN 2150-8097. \\
\href{https://doi.org/\vldbdoi}{doi:\vldbdoi} \\
}\addtocounter{footnote}{-1}\endgroup

\ifdefempty{\vldbavailabilityurl}{}{
\vspace{.3cm}
\begingroup\small\noindent\raggedright\textbf{PVLDB Artifact Availability:}\\
The source code, data, and/or other artifacts have been made available at \url{\vldbavailabilityurl}.
\endgroup
}

\section{Introduction}
\label{sec:intro}

System performance is critical in many database applications where users need answers quickly to gain timely insights and make mission-critical decisions.  In the large body of optimization literature~\cite{conf/icde/FinanceG91, journals/vldb/KossmannPN22, conf/sigmod/PiraheshHH92}, one family of technique is query rewriting, which transforms a query to a new query that computes the same answers with a higher performance. 

\setlength{\columnsep}{5pt}%
\begin{wrapfigure}{r}{0.4\linewidth}
 \centering
    \includegraphics[scale=.75]{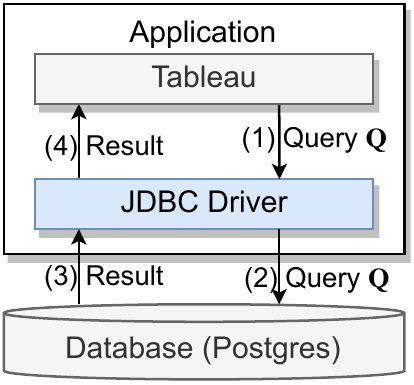}
    \caption{Query lifecycle between Tableau and Postgres.}
  \label{fig:tableau-on-postgres}
\end{wrapfigure}

\boldstart{Motivating example.}
Figure~\ref{fig:tableau-on-postgres} shows a case where a user runs Tableau on top of a Postgres database to analyze and visualize the underlying data of social media tweets.  Tableau formulates and sends a SQL query to the database for each frontend request through a connector such as a JDBC driver.  The database returns the result to Tableau to render in the frontend.

Figure~\ref{fig:original-query-strpos} shows an example SQL query $Q$ formulated by Tableau to compute a choropleth map of tweets containing a substring, e.g., {\tt covid}, which matches {\tt covid-19}, {\tt covid19}, {\tt postcovid}, {\tt covidvaccine}, etc.  Without any index available on the table, the database engine uses a scan-based physical plan, which takes $34$ seconds in our evaluation.  To improve the performance, the developer is tempted to create an index on the \texttt{content} attribute of the table.

\begin{figure}[htbp]
  \centering
  \begin{subfigure}[t]{0.48\linewidth}
    \begin{sqlnotitle}
SELECT SUM(1) AS "cnt:tweets",
                    "state_name" AS "state_name"
FROM "tweets"
WHERE &STRPOS(LOWER("content"),
                    'covid') > 0& 
GROUP BY 2;
    \end{sqlnotitle}
    \caption{An original query $Q$ formulated by Tableau to compute a choropleth map of tweets containing {\em covid} as a substring. \label{fig:original-query-strpos}}
  \end{subfigure}
  \hfill
  \begin{subfigure}[t]{0.48\linewidth}
    \begin{sqlnotitle}
SELECT SUM(1) AS "cnt:tweets",
                    "state_name" AS "state_name"
FROM "tweets"
WHERE &"content" ILIKE 
                    '%covid%'&
GROUP BY 2;
    \end{sqlnotitle}
    \caption{A rewritten query $Q'$ equivalent to $Q$ but runs 100 times faster by using a trigram index on the {\tt content} attribute. \label{fig:rewritten-query-strpos}}
  \end{subfigure}
  \caption{An example query pair (differences shown in blue).}
  \label{fig:rewrite-example-strpos}
  \vspace{-6pt}
\end{figure}

Unfortunately, Postgres does not support an index-based physical plan for the \texttt{STRPOS(LOWER("content"),$s$)} expression in $Q$, where $s$ is an arbitrary string. Interestingly, another query $Q'$, shown in Figure~\ref{fig:rewritten-query-strpos}, is equivalent to $Q$, and uses an \texttt{ILIKE} predicate.  This expression can be answered using a trigram index on the \texttt{content} attribute~\cite{postgres:trigram}, and the corresponding physical plan takes $0.32$ seconds only.  Notice that the optimizer does not produce an index-based plan for the original {\tt STRPOS} predicate using this trigram index~\cite{email:index-in-strpos-function}.

A natural question is whether we can let Tableau generate $Q'$ instead of $Q$ for the database.  Tableau is a proprietary application layer, and has its own internal logic to generate queries, which the developer, in this example, cannot change. We may also consider using the \texttt{CREATE RULE} interface provided by Postgres~\cite{postgres:query-rewrite} to introduce a rewriting rule inside the database, but as we will show in Section~\ref{sec:insufficiency-of-existing-solutions}, this language has limited expressive power and does not allow us to rewrite $Q$ to $Q'$.  As a consequence, we miss the rewriting opportunity to significantly improve the query performance.  \rone{Note that as shown in Section~\ref{sec:end-to-end-query-time}, the rewriting need is not limited to simple predicate levels but also includes complex statement levels.}

\vspace{1mm}
\boldstart{Problem Formulation.}
\rmeta{Besides the above example, as more cases in Section~\ref{sec:insufficiency-of-existing-solutions} and our experiments using different applications and databases on both synthetic and real-world datasets in Section~\ref{sec:end-to-end-query-time} show}, there is a unique problem in a wide range of database-supported systems with the following setting.  (1) {\em The developers need to treat the application layer as a black box and cannot modify its logic of generating SQL queries.}  Reasons include i) the application is proprietary software (e.g., Tableau) and its source code is not available; and ii) the source code of the application is too complicated or old to modify, especially for legacy systems~\cite{conf/sigmod/KhuranaH21}. For example, reports~\cite{blog:software-is-fragile} show that there are many applications where parties have even lost their original source code.   (2) {\em The developers need to treat the database as a black box.}  Reasons include i) the developers do not have the privileges to modify the database; and ii) the database is used by many clients, and the developers want to avoid side effects of database changes to these clients.  (3) {\em The developers want to use their knowledge about the data and domain to rewrite queries sent from the application to the database to significantly improve their performance.}  For example, they may introduce rewriting rules that are valid for their particular database with certain properties (e.g., specific attribute types, certain cardinality constraints, or primary keys and foreign keys), even though these rules may not be valid for all databases.  \rtwo{Specifically, the experimental results in Section~\ref{sec:end-to-end-query-time} illustrate cases where a rewriting is valid only for a particular dataset, and may not be correct in general, thus it cannot be adopted by a database query optimizer.}  Thus, we want to allow developers to be ``in the driver's seat'' during the lifecycle of a query to generate an equivalent and more efficient query as ``human-centered query rewriting''.  \rtwo{Note that we do not seek to replace query optimizers inside databases but only provide a chance for users to inject their knowledge to optimize queries before they are sent to the database.}  Hence, the problem is stated as:

\vspace{1mm}
\noindent\fbox{
    \parbox{0.96\linewidth}{
        \textbf{Problem Statement}: Given an application and a database as black boxes, develop a middleware solution for users to easily express their rules to rewrite application queries for a better performance.
    }
}
\vspace{1mm}

\boldstart{Solution overview.}
In this paper, we propose \sysname, a novel middleware-based service architecture for human-centered query rewriting.  It is between an application and a database, intercepts SQL queries sent from the application, and rewrites them using human-crafted rewriting rules to improve their performance.  By providing a slightly-modified JDBC/ODBC driver or a RESTful proxy for the query interception, \sysname requires no code changes to either the application or the database.  \sysname provides an expressive and easy-to-use rule language (called \langname) for users \rone{(SQL developers or DBAs)} to define rewriting rules \rone{(i.e., customizing query rewriting for their application queries)}.  Users can easily express their rewriting needs by providing the query pattern and its replacement.  They can also specify additional constraints and actions for complex rewriting details.  In addition, \sysname allows users to express their rewriting intentions by providing examples.  That is, users can input original queries and the desired rewritten queries.  Then \sysname automatically generalizes the examples into rewriting rules and suggests high-quality rules.  The users can confirm the rules to be saved in the system or further modify the rules as they want.  

\vspace{1mm}
\boldstart{Challenges and contributions.}
To develop the \sysname system, we face several challenges. {\em (C1)} How to develop an expressive and easy-to-use rule language for users to formulate rules?  {\em (C2)} How to generalize pairs of original and rewritten queries to rewriting rules and measure their quality?  {\em (C3)} How to search the candidate rewriting rules to suggest high-quality ones based on the user-given examples?  In this paper, we study these challenges and make the following contributions.

\begin{itemize}
    \item We propose a novel middleware-based query rewriting service to fulfill the need of human-centered query rewriting (Section~\ref{sec:querybooster-middleware}).
    \item We study the suitability of existing rule languages in the literature and show their limitations.  We then develop a novel rule language (\langname) that is expressive and easy to use (Section~\ref{sec:rule-lang}).
    \item We develop transformations to generalize pairs of rewriting queries to rules and propose using the metric of minimum description length to measure rule quality (Section~\ref{sec:rule-qualities-and-transformations}).
    \item We present a framework to search the candidate rewriting rules efficiently and suggest high-quality rules based on user-given examples (Section~\ref{sec:searching-high-quality-rules}).
    \item We conduct a thorough experimental evaluation, including a user study, to show the benefits of the \langname rule language, the effectiveness of the rule-suggesting framework, and the advantages of human-centered query rewriting (Section~\ref{sec:experiments}).
\end{itemize}

\section{Related Work and Limitations 
\label{sec:insufficiency-of-existing-solutions}}

\begin{figure}[htb]
  \centering
  \includegraphics[width=.99\linewidth]{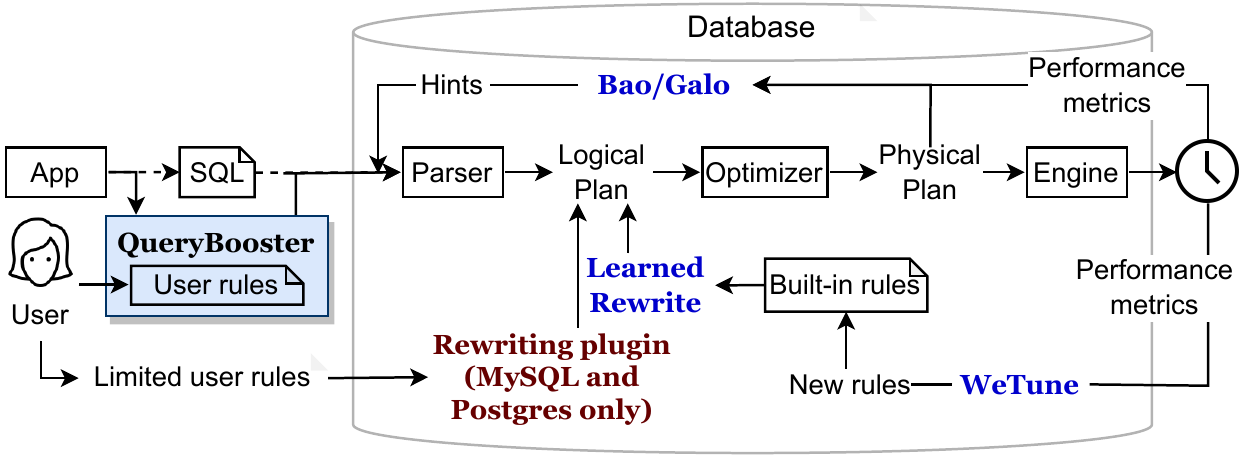}
  \caption{Query-rewriting solutions for databases (native solutions in brown and third-party solutions in blue).}
  \label{fig:query-rewriting-systems}
  \vspace{-2pt}
\end{figure}

In this section, we show existing solutions and why they cannot solve the formulated problem.  Figure~\ref{fig:query-rewriting-systems} gives an overview of various solutions for query rewriting in the lifecycle of a query in a database system~\cite{calcite, journals/pvldb/DamasioCGMMSZ19, conf/sigmod/MarcusNMTAK21, conf/sigmod/WangZYDHDT0022, journals/pvldb/ZhouLCF21} and the position of the proposed \sysname system.  At a high level, these solutions can be classified into two categories: native writing plugins and third-party solutions.

\boldstart{Native rewriting plugins.}  Most databases such as AsterixDB~\cite{asterix:website}, IBM DB2~\cite{db2:doc}, MongoDB~\cite{mongodb:doc}, MS SQL Server~\cite{sqlserver:doc}, MySQL~\cite{mysql:query-rewrite}, Oracle~\cite{oracle:query-rewrite}, Postgres~\cite{postgres:query-rewrite}, SAP HANA~\cite{hana:doc},  Snowflake~\cite{snowflake:website}, and Teradata~\cite{teradata:website}, do not have capabilities for users to rewrite queries sent to the database.  Notice that even though ``hints'' can be included in a query to make suggestions to the database optimizer, they are technically not used to change the query, thus, are not a query-rewriting solution.  To our best knowledge, only two database systems, Postgres and MySQL, provide a plugin for users to define new rules to rewrite queries before sending them to the database.  However, their rule-definition languages have limited expressive power, as discussed below.  

\textbf{\textit{Postgres.}} A rewriting rule in the Postgres plugin can only define a pattern matching a table name in a {\sf SELECT} clause of a SQL query and replace the table with another table or a subquery~\cite{postgres:query-rewrite}.  Its rule language cannot express the rewriting in the running example in Figure~\ref{fig:rewrite-example-strpos}.  In particular, it does not support a pattern that matches a component in a SQL statement at the predicate level, e.g., the \texttt{STRPOS(LOWER("content"), \_) > 0} portion in the {\tt WHERE} clause in the original query.  Safety could be a major consideration behind this rule language.  For instance, the Postgres 14 documentation~\cite{postgres14:doc} explained that ``{\em this restriction was required to make rules safe enough to open them for ordinary users, and it restricts ON SELECT rules to act like views}.''

\textbf{\textit{MySQL.}}  The MySQL plugin uses the syntax of prepared statements to define query-rewriting rules, and a rule replaces a SQL query matching the rule's pattern with a new statement~\cite{mysql:query-rewrite}.  A rule includes placeholders that can only match literal values in a SQL query, such as a constant in a predicate in the {\tt WHERE} clause. A main limitation of this language is that a placeholder cannot match many components in a query, such as table names and attribute names.  For instance, the following is a predicate from a query formulated by Tableau to MySQL:
\begin{rule-example}
adddate(date_format(`created_at`, '$\%$Y-$\%$m-$01$ $00$:$00$:$00$'), 
                   interval $0$ second) = TIMESTAMP('$2018$-$04$-$01$ $00$:$00$:$00$')
\end{rule-example}
And if we rewrite the predicate by removing the type-casting on the right-hand constant, as shown below:
\begin{rule-example}
adddate(date_format('created_at', '$\%$Y-$\%$m-$01$ $00$:$00$:$00$'), 
                   interval $0$ second) = '$2018$-$04$-$01$ $00$:$00$:$00$'
\end{rule-example}
The corresponding rewritten query is significantly faster ($2.68$s) than the original query ($87$s).  Unfortunately, the MySQL plugin does not support this rewriting because a pattern in the MySQL plugin has to be an {\em entire} statement instead of a {\em single} predicate. In other words, using the MySQL plugin for this rewriting requires the enumeration of all other parts of the target SQL query.

%\rone{In summary, \sysname provides a unique and powerful interface for users to define rewriting rules that are not supported by Postgres or MySQL's user-facing rewriting plugins.}

\boldstart{Third-party solutions.}  Bao~\cite{conf/sigmod/MarcusNMTAK21} and Galo~\cite{journals/pvldb/DamasioCGMMSZ19} rewrite queries by adding hints to help the database optimizer generate more efficient physical plans based on their cost estimations and searching methods.  They take a physical plan and query performance as the input and produce hints to the original query.  WeTune~\cite{conf/sigmod/WangZYDHDT0022} generates new rewriting rules automatically by searching the logical-plan space and considering the performance of rewritten queries.  LearnedRewrite~\cite{journals/pvldb/ZhouLCF21} utilizes built-in rewriting rules inside the database to optimize queries, and the users have no control over when and which rules are applied. 
None of these solutions allow users to formulate their own rewriting rules to fulfill the human-centered query rewriting need.  
\rmeta{PgCuckoo~\cite{conf/sigmod/HirnG19} opens an opportunity for users to inject intelligent logic to manipulate query plans in Postgres.  It only works for Postgres and the proposed middleware solution works for any databases.}

\boldstart{Commercial systems.} There are also commercial systems that do query rewriting for applications on top of databases.  For example, Keebo~\cite{keebo:website} uses data learning and {\em approximate} query processing (AQP) techniques to accelerate analytical queries.  It runs queries on summarized tables instead of the raw data as much as possible to reduce query time.  EverSQL~\cite{eversql:website} uses AI/ML techniques to recommend rewriting ideas for queries on MySQL and Postgres.  Other systems such as ApexSQL~\cite{apexsql:website}, Query Performance Insights for Azure SQL~\cite{query-insight-azure-sql:website}, and Toad~\cite{toad:website} help database developers analyze query performance bottlenecks and tune database knobs. None of these systems allow users to formulate their own rewriting rules to fulfill the human-centered query rewriting need.

\rfour{
\boldstart{General pattern-matching tools.}  These tools can be used to rewrite any program and are not limited to SQL code.  For instance, Quasiquotation~\cite{journals/pacmpl/ParreauxVSK18, website:scala-quasiquotes} is a general technique to rewrite programs using meta-programs.  A main issue of the tools is that they are not designed for SQL queries, and they do not consider the unique semantics (tables, columns, etc.) of SQL, which is considered by the proposed \sysname.
}

%TODO:

%\boldstart{Uniqueness of \sysname.}  To our best knowledge, the proposed system is the only solution that allows users to define their own rules to rewrite queries before the queries are sent to the database.  In this way, it can not only provide a chance for the users to add rules based on their domain knowledge but also allow minimal or no changes to the application and database. 

\section{\sysname: Overview \label{sec:querybooster-middleware}}

In this section, we present a novel middleware system called \sysname, to fulfill the need for human-centered query rewriting.  We first show its architecture and then discuss the two tasks in developing the system.

%\subsection{Architecture}

\begin{figure}[htb]
  \centering
  \includegraphics[width=0.99\linewidth]{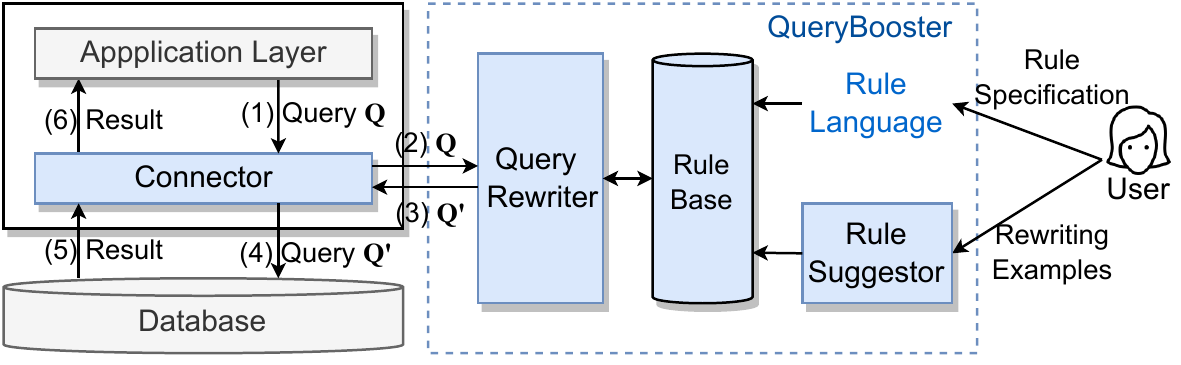}
  \caption{Architecture of \sysname.}
  \label{fig:architecture}
  \vspace{-6pt}
\end{figure}

Figure~\ref{fig:architecture} shows the architecture of \sysname.  %It includes two components, namely  {\em query rewriting} and {\em rule formulation}.
\rone{It includes two phases, an offline {\em rule formulation} phase and an online {\em query rewriting} phase.}

\rone{For the offline {\em rule formulation} phase, }
\sysname provides a powerful interface for users to formulate rewriting rules.  It allows users to formulate rules in the following two ways.  First, it provides an expressive and easy-to-use rule language for users to define rewriting rules.  Users can easily express their rewriting needs by writing down the query pattern and its replacement.  They can also specify additional constraints and actions to express complex rewriting details.  Second, it allows users to express their rewriting intentions by providing examples.  A rewriting example is a pair of SQL queries with the original query and the desired rewritten query.  The ``Rule Suggestor'' automatically suggests high-quality rewriting rules based on the examples.  The users can choose their desired rewriting rules and further modify suggested rules as they want.  \rone{All user-confirmed rules are stored in the ``Rule Base,'' and the ``Query Rewriter'' will rewrite online queries based on the rules.}
%To implement such a powerful interface, we have the following two tasks.

\rone{For the online {\em query rewriting} phase, } %For query rewriting, 
\sysname provides a customized connector that communicates with its service to rewrite application queries.  In particular, the connector accepts an original query $Q$ formulated by the application and sends $Q$ to the ``Query Rewriter'' service, which applies rewriting rules stored in the ``Rule Base'' to rewrite $Q$ to a new query $Q'$.  The new query is sent back to the connector, which forwards $Q'$ to the backend database to boost the application's performance.  \rone{Note that \sysname focuses on rewriting queries based on user-specified rules and assumes no access to the backend database to create indexes.}

%For example, Figure~\ref{fig:original-query-strpos} shows an original SQL query $Q$ formulated by an application (e.g., Tableau), and Figure~\ref{fig:rewritten-query-strpos} shows a rewritten query $Q'$ desired by the user who wants to apply a rewriting rule on $Q$ to replace the ``\texttt{STRPOS}'' function with the ``\texttt{ILIKE}'' predicate.  In this case, the query rewriter inputs query $Q$, applies the user-defined rule on $Q$, rewrites it into $Q'$, and outputs $Q'$.

\rone{To use the \sysname rewriting service, users do not need to modify any code of the applications or databases or install any plugins. They only need to replace the existing DB connector with a \sysname-customized one.}
The connector can be for either an ODBC/JDBC interface or a RESTful interface.  Most database vendors provide ODBC/JDBC drivers with an open-source license.  Thus we can provide a slightly modified version of the driver that communicates with the proposed \sysname service to rewrite queries for these databases.  For instance, in our developed prototype~\cite{conf/edbt/BaiA022}, we added only 112 lines of code to the PostgreSQL JDBC driver.  For databases with redistribution restrictions on their drivers (e.g., Oracle JDBC driver~\cite{oracle:free-license}), we can provide users a software patch with a small amount of source code modifications.  For applications and databases that communicate through a RESTful interface, we can provide a proxy web server that forwards all requests and responses between them transparently.  The proxy server in the middle rewrites an application request by communicating with the Query Rewriter service.  We assume the RESTful API endpoint in the application is configurable, i.e., we can switch the target database endpoint to our service.  To develop \sysname, we have the following two tasks.

\boldstart{(Task 1) Developing an expressive and easy-to-use rule language.}  The first task of \sysname is to provide an expressive and easy-to-use rule language for users to formulate rules.  It should meet the following three requirements.  {\em (R1) Powerful expressiveness in SQL semantics.}  It needs to understand SQL-specific semantics where users can specify pattern-matching conditions on the elements of a SQL query, e.g., two tables have the same name, or a column is unique in the table schema.  {\em (R2) Easy to use by SQL users.}  Users of \sysname are application developers who are familiar with SQL. \sysname should require users to have little prior knowledge other than SQL to define their rewriting rules.  {\em (R3) Independent of databases or SQL dialects.}  As a general query-rewriting service, the rule language should be independent of any specific database or SQL dialect.  Providing a rule language that meets the aforementioned three requirements is challenging.  In Section~\ref{sec:rule-lang}  we first study the suitability of existing rule languages in the literature and then develop a novel rule language that combines advantages from existing languages to meet all the requirements. 

\boldstart{(Task 2) Suggesting rules from examples.}  The second task of \sysname is to provide a rule-suggestion framework that automatically generalizes user-given rewriting examples into rewriting rules and suggests high-quality ones.  Manually formulating a rewriting rule that covers many queries is tedious.  We want a better experience in which a user expresses the rewriting intention by providing query rewriting pairs.  Then, the system can automatically suggest rewriting rules to achieve the rewritings of the given examples.  For instance, if a user inputs the rewriting pair in Figure~\ref{fig:rewrite-example-strpos} as one of the examples, the rule suggestor can automatically generalize it and recommend the following rule to the user.

\begin{rule-example}
STRPOS(LOWER(<x>), '<y>') --> <x> ILIKE '%<y>%'
\end{rule-example}

Developing such a rule suggestor is also challenging since we need to answer a few questions, such as how to measure the quality of rewriting rules, how to generalize query rewriting pairs into rewriting rules, and how to search the candidate rewriting rules to suggest high-quality ones.  We answer them in Sections~\ref{sec:rule-qualities-and-transformations} and ~\ref{sec:searching-high-quality-rules}.

\rfour{
\boldstart{Correctness of rewriting rules.}  In the case where users make mistakes when formulating rewriting rules, we can leverage existing query equivalence verifiers (e.g., ~\cite{journals/pvldb/ChuMRCS18, conf/sigmod/WangZYDHDT0022}) to validate the rules and guarantee their correctness. 
}

\section{\langname: A Rewriting-Rule Language}
\label{sec:rule-lang}

In this section, we focus on providing an expressive and easy-to-use rule language for \sysname's users to formulate rewriting rules that work for different applications and databases.  We first study the suitability of existing rule languages in the literature and then develop a novel rule language that meets all the requirements desired by \sysname.

\subsection{Suitability of Existing Rule Languages}

\begin{figure}[htb]
  \centering
  \includegraphics[width=.95\linewidth]{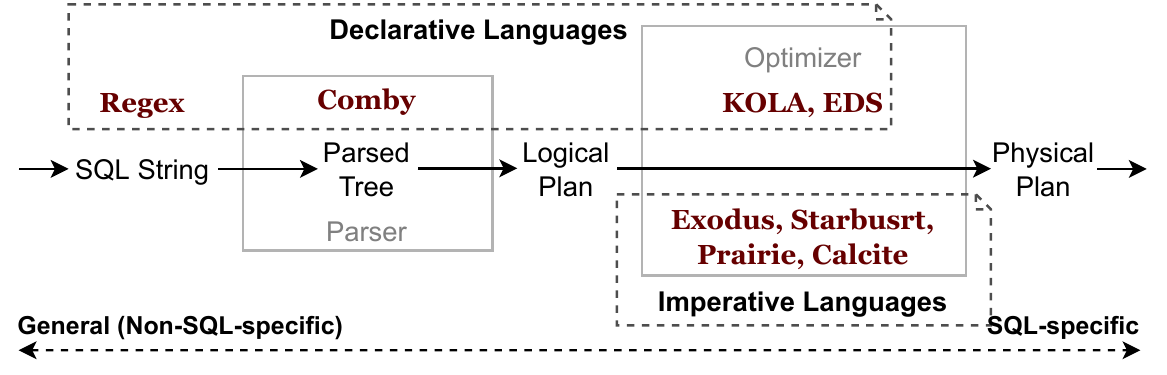}
  \caption{Existing rule languages (shown in brown) in the lifecycle of a SQL query.}
  \label{fig:rule-langs}
  \vspace{-2pt}
\end{figure}

We study existing rule languages and their suitability for \sysname.  These languages~\cite{conf/sigmod/GraefeD87, conf/icde/ScioreS90, conf/icde/FinanceG91, conf/sigmod/PiraheshHH92, conf/icde/DasB95, conf/sigmod/CherniackZ96, conf/icde/PiraheskLH97, wiki:regular-expression, comby:website, calcite} are summarized in Figure~\ref{fig:rule-langs}. 
We categorize the languages in two aspects: general versus SQL-specific and declarative versus imperative, and then summarize how the languages in different categories meet the requirements of the rule language of \sysname.

\boldstart{General versus SQL-specific.}  The languages on the left are more general (i.e., non-SQL-specific) since they have fewer SQL-specific restrictions.  For example, regular expressions~\cite{wiki:regular-expression} (shown as ``Regex'') do not require the input string to be a SQL query.  On the contrary, rule languages on the right such as EDS~\cite{conf/icde/FinanceG91} only accept valid SQL query plans as the input.  

\begin{figure}[htbp]
  \footnotesize
  \centering
  \begin{subfigure}[t]{0.48\linewidth}
    \centering
    \includegraphics[scale=0.65]{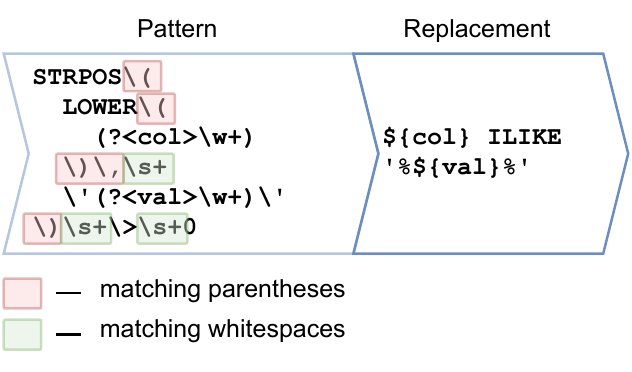}
    \caption{A rule written in regex.}
    \label{fig:regex-rule-strpos}
  \end{subfigure}
  \hfill
  \begin{subfigure}[t]{0.48\linewidth}
    \centering
    \includegraphics[scale=0.65]{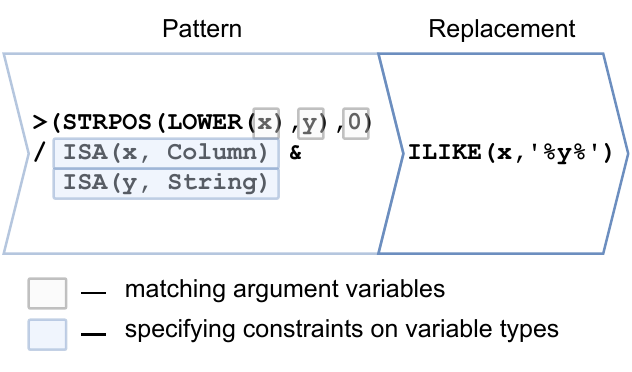}
    \caption{A rule written in the EDS language~\cite{conf/icde/FinanceG91}.}
    \label{fig:eds-rule-strpos}
  \end{subfigure}
  \caption{Rewriting rules to replace \texttt{STRPOS} functions with \texttt{ILIKE} predicates in two different languages.}
  \label{fig:rules-strpos}
  \vspace{-2pt}
\end{figure}

The main advantage of SQL-specific languages is that they are very powerful for users to express rewriting rules in SQL-specific semantics.  For instance, to achieve the rewriting of replacing ``\texttt{STRPOS}'' functions with ``\texttt{ILIKE}'' predicates, we can write a rule using either regex as shown in Figure~\ref{fig:regex-rule-strpos}, or EDS as shown in Figure~\ref{fig:eds-rule-strpos}.  Compared to regex, the EDS language has two advantages.  First, we do not need to specify SQL-specific syntax requirements such as white spaces and arguments in functions.  Second, we can specify SQL-specific constraints for variables that are not supported by regex, such as ``\texttt{x} is a column and \texttt{y} is a \texttt{String} literal.'' 

% The main advantage of SQL-specific languages is that they can utilize SQL-specific syntax and semantics to help users formulate complex rewriting rules.  For instance, to achieve the rewriting of replacing ``\texttt{STRPOS}'' functions with ``\texttt{ILIKE}'' predicates, we can write a rule using either regex as shown in Figure~\ref{fig:regex-rule-strpos}, or EDS as shown in Figure~\ref{fig:eds-rule-strpos}.  Compared to regex, the EDS language has two advantages.  First, we do not need to specify SQL-specific syntax requirements such as white spaces and arguments in functions.  Second, we can specify SQL-specific constraints for variables that are not supported by regex, such as ``\texttt{x} is a column and \texttt{y} is a \texttt{String} literal."  

There are also disadvantages of the more specific languages. First, they can be limited to a particular SQL dialect or database. For instance, EDS is designed for a particular extensible system (called ``EDBMS'')  and its SQL dialect~\cite{conf/icde/FinanceG91}.  Second, they require the users to deeply understand how a SQL query is translated into a plan and how the database optimizer works.  For example, to formulate the rule in the EDS language shown in Figure~\ref{fig:eds-rule-strpos}, a user has to translate the original SQL predicate of ``\texttt{STRPOS($\cdots$) > 0}'' into a logical plan tree format with ``\texttt{>}'' as the parent node, which can be counter-intuitive for end users who are familiar with the SQL syntax but not a database engine.  

% Similarly, KOLA~\cite{conf/sigmod/CherniackZ96} is an algebraic language designed to implement a query optimizer inside a database.  \qiushi{It requires users prior knowledge of not only how query optimization works but also how combinator-based algebra works for expressing rewriting logic.}

% To this end, \sysname desires a rule language with the following requirements. \qiushi{First, \textbf{(r1) a user familiar with SQL can use it.} Second, \textbf{(r2) it understands SQL-specific syntax and semantics} such as delimiters, keywords, tables, columns, etc.  Third, since \sysname is a general middleware system serving different applications and databases, \textbf{(r3) the language should support various databases and SQL dialects.}}

\boldstart{Declarative versus Imperative.} The existing rule languages are either {\em declarative} or {\em imperative}.  In a declarative rule language, users describe how a rule (e.g., pattern and replacement) looks like, but not how a rule should be implemented.  On the contrary, in an imperative rule language, users specify a sequence of steps that should be taken to do the rewriting.  For example, regex is declarative. Calcite~\cite{calcite} is imperative, and it uses Java to formulate a rule.  The primary disadvantage of using an imperative language to define rules is that it requires users to have prior knowledge about the internal structures of the rule engine and define rules by writing code.  For example, in Calcite, a user has to write a Java class that implements an interface to define a new rule.  

The main advantage of imperative languages is the expressive power offered by the programming language (e.g., C++), such as defining schema-dependent pattern-matching conditions.  For example, a rewriting rule that removes unnecessary self-joins may need to verify the joining attribute is unique, which cannot be inferred just from the SQL query itself.  Thus, we need schema information from the database to implement this rule.  Using an imperative rule engine, we can easily write a rule with a few lines of C code~\cite{conf/icde/PiraheskLH97} that accesses the schema data and checks the matching condition.  

To this end, we summarize how existing languages meet \sysname's requirements on its rule language in Table~\ref{tab:rule-langs}.  An observation is that no existing rule language satisfies all the requirements.  Next, we develop a novel rewriting-rule language called \langname. 

\begin{table}[htbp]
\footnotesize
\caption{Suitability of existing languages for \sysname.}
\label{tab:rule-langs}
\begin{tabular}{|l|ll|l|l|}
\hline
\multirow{2}{*}{\textbf{\begin{tabular}[c]{@{}l@{}}Rule\\ Language\end{tabular}}} & \multicolumn{2}{l|}{\textbf{Expressive Power}} & \multirow{2}{*}{\textbf{\begin{tabular}[c]{@{}l@{}}Indepen-\\dent of\\ DB\end{tabular}}} & \multirow{2}{*}{\textbf{\begin{tabular}[c]{@{}l@{}}Additional \\ Knowledge \\ Users Need\end{tabular}}} \\ \cline{2-3}
 & \multicolumn{1}{l|}{\textit{\begin{tabular}[c]{@{}l@{}}SQL \\ Semantics\end{tabular}}} & \textit{\begin{tabular}[c]{@{}l@{}}SQL \\ Schema\end{tabular}} &  & \\ \hline
\textbf{Regex} & \multicolumn{1}{l|}{No} & No & \textbf{Yes} &  \\ \hline
\textbf{Comby} & \multicolumn{1}{l|}{No} & No & \textbf{Yes} &  \\ \hline
\textbf{KOLA} & \multicolumn{1}{l|}{\textbf{Yes}} & No & \textbf{Yes} & \begin{tabular}[c]{@{}l@{}}\circled{1} \circled{2} \circled{3}\end{tabular} \\ \hline
\textbf{EDS} & \multicolumn{1}{l|}{\textbf{Yes}} & \textbf{Yes} & No & \begin{tabular}[c]{@{}l@{}}\circled{1} \circled{2}\end{tabular} \\ \hline
\textbf{Exodus} & \multicolumn{1}{l|}{\textbf{Yes}} & \textbf{Yes} & No & \begin{tabular}[c]{@{}l@{}}\circled{1} \circled{2}\end{tabular} \\ \hline
\textbf{Starburst} & \multicolumn{1}{l|}{\textbf{Yes}} & \textbf{Yes} & No & \begin{tabular}[c]{@{}l@{}}\circled{1} \circled{2} \circled{4} \circled{5}\end{tabular} \\ \hline
\textbf{Prairie} & \multicolumn{1}{l|}{\textbf{Yes}} & \textbf{Yes} & No & \begin{tabular}[c]{@{}l@{}}\circled{1} \circled{2} \circled{5}\end{tabular} \\ \hline
\textbf{Calcite} & \multicolumn{1}{l|}{\textbf{Yes}} & \textbf{Yes} & \textbf{Yes} & \begin{tabular}[c]{@{}l@{}}\circled{1} \circled{2} \circled{4} \circled{6}\end{tabular} \\ \hline
\textbf{VarSQL} & \multicolumn{1}{l|}{\textbf{Yes}} & \textbf{Yes} & \textbf{Yes} &  \\ \hline
\end{tabular}
  \\ 
  \circled{1} Query Optimization; \circled{2} Relational Algebra; \circled{3} Combinator-based Algebra; 
  \\
  \circled{4} Internal Data Structure; \circled{5} C++ Programming; \circled{6} Java Programming;
\vspace{-2pt}
\end{table}

\subsection{\langname: A Novel Rule Language}

We develop a novel rewriting-rule language (called \langname\footnote[1]{\langname stands for ``Variablized SQL''.}) for \sysname that meets all the requirements.
%to provide an expressive and easy-to-use interface for users to define rewriting rules that work for different applications and databases.
%\qiushi{As far as we know, \langname is the first rewriting-rule language for SQL queries that meets all the requirements asked by \sysname.   
In particular, \langname understands SQL-specific semantics and supports schema-dependent pattern-matching conditions (R1).  It is easy to use, requiring no prior knowledge other than SQL (R2).  Also, it is independent of any specific database or SQL dialect (R3).  Next, we present the technical details of \langname.

%\boldstart{Annotating SQL queries with variables}.  The main idea of \langname is to annotate existing elements (e.g., tables, columns, values, expressions, predicates, sub-queries, etc.) in SQL queries as variables in a rule's pattern and replacement parts such that the annotated SQL query can pattern-match a broad set of SQL queries and manipulate the replacement arbitrarily.  We call this annotating process ``variablizing'' a SQL query and the new SQL query a ``variablized'' SQL query.  \langname is unique in that it annotates SQL elements as variables (similar to SQL-specific languages such as EDS) and defines a rule pattern with a plain SQL-query string (similar to general languages such as Comby).

The syntax of \langname to define a rewriting rule is as follows:

\begin{rule-example}
[Rule] ::= [Pattern] / [Constraints] --> [Replacement] / [Actions].
\end{rule-example}

\langname uses a four-component structure adopted by most rule languages (e.g., EDS~\cite{conf/icde/FinanceG91} and Comby~\cite{comby:website}). The ``Pattern'' and ``Replacement'' components define how a query is matched and rewritten into a new query.  The ``Constraints'' component defines additional conditions that cannot be specified by a pattern such as schema-dependent conditions.  The ``Actions'' component defines extra operations that the replacement cannot express, such as replacing a table's references with another table's.  We first discuss how a pattern and a replacement are formulated using \langname.

\boldstart{Extending SQL with variables}.  The main idea of using \langname to define a rule's pattern is to extend the SQL language with variables. A variable in a SQL query pattern can represent an existing SQL element such as a table, a column, a value, an expression, a predicate, a sub-query, etc.  In this way, a user can formulate a query pattern as easily as writing a normal SQL query.  The only difference is that, using \langname, one can use a variable to represent a specific SQL element so that the pattern can match a broad set of SQL queries.  We call this pattern-formulating process ``variablizing'' a SQL query, and we call the formulated pattern query a ``variablized'' SQL query.  Similarly, a rule's replacement is formulated by writing the rewritten SQL query using variables introduced in the rule's pattern.  Particularly, both the pattern and replacement in a \langname rule have to be a full or partial SQL query optionally variablized.  The variables and their matching conditions are defined in Table~\ref{tab:rule-variables}. 

%\langname is unique in that it represents SQL elements as variables (similar to SQL-specific languages such as EDS) and defines a rule pattern with a plain SQL-query string (similar to general languages such as Comby).

\begin{table}[htbp]
\footnotesize
\caption{Variable definitions in \langname.}
\label{tab:rule-variables}
\begin{tabular}{|l|l|l|l|}
\hline
\textbf{Name} & \textbf{Syntax (regex)} & \textbf{Description} & \textbf{Example} \\ \hline
\begin{tabular}[c]{@{}l@{}}Element-\\Variable\end{tabular} & <[a-zA-Z0-9\_]*> & \begin{tabular}[c]{@{}l@{}}An element-variable \\ matches  \\ a table, \\ a column, \\ a value, \\ an expression, \\ a predicate, \\ or a sub-query.\end{tabular} & \begin{tabular}[c]{@{}l@{}}\texttt{STRPOS(LOWER(<x>),} \\ \texttt{'iphone') > 0}\\ \\ \texttt{<x>} matches any \\ value, \\ column, \\ expression, \\ or sub-query.\end{tabular} \\ \hline
\begin{tabular}[c]{@{}l@{}} Set-\\Variable\end{tabular} & <{}<[a-zA-Z0-9\_]*>{}> & \begin{tabular}[c]{@{}l@{}}A set-variable \\ matches \\ a set of \\ tables, \\ columns, \\ values, \\ expressions,\\ predicates, \\ or sub-queries.\end{tabular} & \begin{tabular}[c]{@{}l@{}}\texttt{SELECT <{}<x>{}>} \\    \texttt{FROM <t>} \\ \texttt{WHERE <{}<p>{}>}\\ \\ <{}<x>{}> matches any \\ set of values, \\ columns, \\ expressions, \\ or sub-queries.\end{tabular} \\ \hline
\end{tabular}
\end{table}

To minimize the learning cost for end-users to define rules in \langname, we introduce only two variables, namely ``element-variable'' and ``set-variable.''  An element-variable can match any individual element in a SQL query, such as a table, a column, etc.  A set-variable can match any collection of elements in a SQL query, such as the column list in the \texttt{SELECT} clause.  Note that keywords and delimiters cannot be represented as variables. An entire clause cannot be represented as any type of variable either.
 For example, a set-variable can match all the columns in the \texttt{SELECT} clause, but no type of variable can match the entire \texttt{SELECT} clause.

 \begin{figure}[htb]
    \centering
    \includegraphics[width=.99\linewidth]{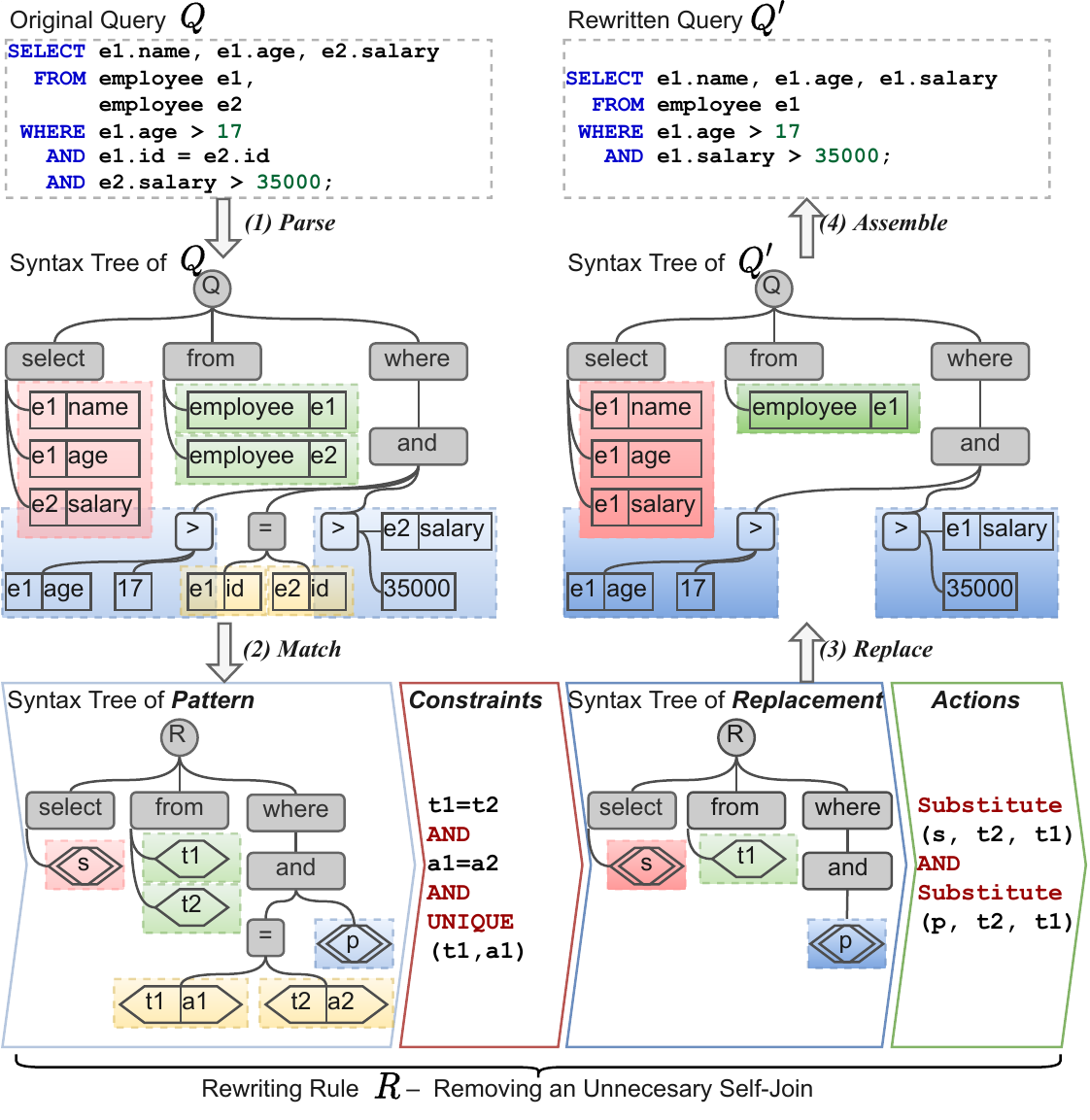}
    \caption{The process of pattern matching and replacing of a \langname rule $R$ on an example query $Q$.  The gray nodes in both syntax trees of $Q$ and the $R$'s pattern are matched keywords.  The colored dashed boxes show the variables in $R$'s pattern and their matched elements in $Q$.}
    \label{fig:rewrite-process}
    \vspace{-2pt}
\end{figure}

\begin{figure}[htb]
    \centering
    \includegraphics[width=.95\linewidth]{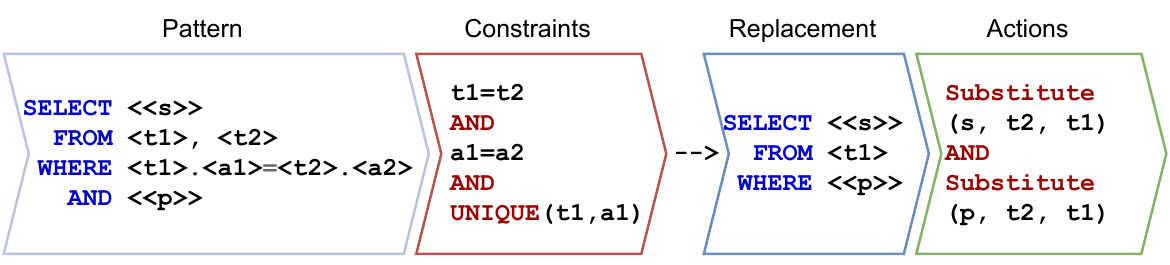}
    \caption{An example rule in \langname that removes an unnecessary self-join.}
    \label{fig:rule-remove-self-join}
    \vspace{-2pt}
\end{figure}

\boldstart{SQL syntax tree-based pattern matching and replacement.}  \langname does the pattern matching and replacement at the SQL syntax tree level.   Consider the rule $R$ shown at the bottom of Figure~\ref{fig:rewrite-process}, where the pattern and replacement are shown in their syntax tree formats.  This rule specifies that when two tables with the same name join on the same unique column, we can safely remove the join and keep only one copy of the table.  The remaining of Figure~\ref{fig:rewrite-process} shows the process of pattern matching and replacement of this rule on an example query $Q$.  We first obtain the syntax tree of the query, and compare it node by node against the syntax tree of the rule's pattern.  The keyword nodes match each other, and the variables in the pattern match those elements in the query.  Under the node ``\texttt{and}'', the subtree ``\texttt{<t1>.<a1>=<t2>.<a2>}'' in the pattern matches the predicate ``\texttt{e1.id=e2.id}'' in the query, and the set-variable ``\texttt{<{}<p>{}>}'' matches the two remaining predicates in the query.  Next, we use the rule's replacement syntax tree as a template to generate the rewritten query's syntax tree by replacing the variables with their matched elements in the pattern.  Finally, we assemble the rewritten query from the syntax tree.  For completeness, we also show the string representation of the rule $R$ in Figure~\ref{fig:rule-remove-self-join}.

\boldstart{Providing pre-implemented imperative procedures.}
% The ``Constraints'' component in a rule is where a user can specify predicates on variables annotated in the pattern of the rule.  For example, the rule in Figure~\ref{fig:rule-remove-self-join} has three constraints that need to be satisfied jointly for the rule's pattern to match a query.  The condition ``\texttt{t1=t2}'' and ``\texttt{a1=a2}'' indicate that both table names in the \texttt{FROM} clause should be the same and both column names in the join predicate ``\texttt{<t1>.<a1>=<t2>.<a2>}'' should be the same as well.  
Based on SQL, \langname is a declarative language.  One problem with declarative languages is that they lack the expressive power to define complex logic in the replacement of a rule and schema-dependent pattern-matching conditions where imperative programs are needed to access the database schema.  To solve this issue, \langname adopts the idea used in declarative languages such as EDS and Comby that it provides pre-implemented imperative procedures for users to define complex logic in the constraint and action components of rules.
For example, the last constraint ``\texttt{UNIQUE(t1, a1)}'' defined in the ``Constraints'' component in the rule shown in Figure~\ref{fig:rule-remove-self-join} calls the pre-implemented imperative procedure ``\texttt{UNIQUE}'' supported by \langname, which verifies if ``\texttt{a1}'' in table ``\texttt{t1}'' is a unique column by referring to the database schema.

\langname also provides imperative procedures for users to define complex actions in a rule.  
% The ``Actions'' component in a rule allows users to manipulate the rewritten result by specifying actions taken on the replacement SQL query.  
For example, the rule in Figure~\ref{fig:rule-remove-self-join} does two actions on the replacement SQL query.  The first action ``\texttt{Substitute(s, t2, t1)}'' is to replace the table ``\texttt{t2}'' with table ``\texttt{t1}'' in the scope represented by the set-variable ``\texttt{<{}<s>{}>}''.  Consider the query $Q$ in Figure~\ref{fig:rewrite-process} that matches the rule.  The set-variable ``\texttt{<{}<s>{}>}'' matches the entire selection list ``\texttt{e1.name, e1.age, e2.salary}''.  Since the replacement of the rule removes table ``\texttt{t2}'' from the query, the column ``\texttt{e2.salary}'' needs to be substituted by ``\texttt{e1.salary}''.  And, the action ``\texttt{Substitute(s, t2, t1)}'' achieves this purpose.  

\rfour{To make sure the pattern-matching and replacement at the syntax tree level can handle SQL semantics, \langname understands important SQL concepts, e.g., an element-variable ``\texttt{<x>}'' in the \texttt{FROM} clause can match either a table name or a table name with an alias.}

%\rfour{The above substitution example also illustrates the restrictions of doing pattern-matching at the syntax tree level.  The substitution actions may not be needed if we do the pattern-matching at a logical plan tree level, where the ownership of attributes and aliases are captured in the tree.  However, one disadvantage of operating at the logical plan tree level is it requires the user to understand how a SQL string is translated into relational algebra.  We choose the syntax tree level mainly due to its simplicity and convenience.}

% Similarly, the second action ``\texttt{Substitute(p, t2, t1)}'' replaces ``\texttt{e2.salary}'' in the \texttt{WHERE} clause with ``\texttt{e1.salary}'', which makes sure \qiushi{the table aliases in} the rewritten query \qiushi{are correct}.

\section{Rule Quality and Transformations} 
\label{sec:rule-qualities-and-transformations}

In this section, we focus on providing a powerful interface for \sysname that suggests high-quality rules for user-given rewriting examples.  We first discuss how to measure the quality of rules and formally define the rewriting-rule suggestion problem.  We then propose a framework to solve the problem, which comprises two major steps: transforming rules into more general forms and searching for high-quality rules greedily.  We discuss the first step in this section and the second step in the next section.

\subsection{Quality of Rewriting Rules}

\begin{figure*}[t]
  \centering
  \includegraphics[width=.95\linewidth]{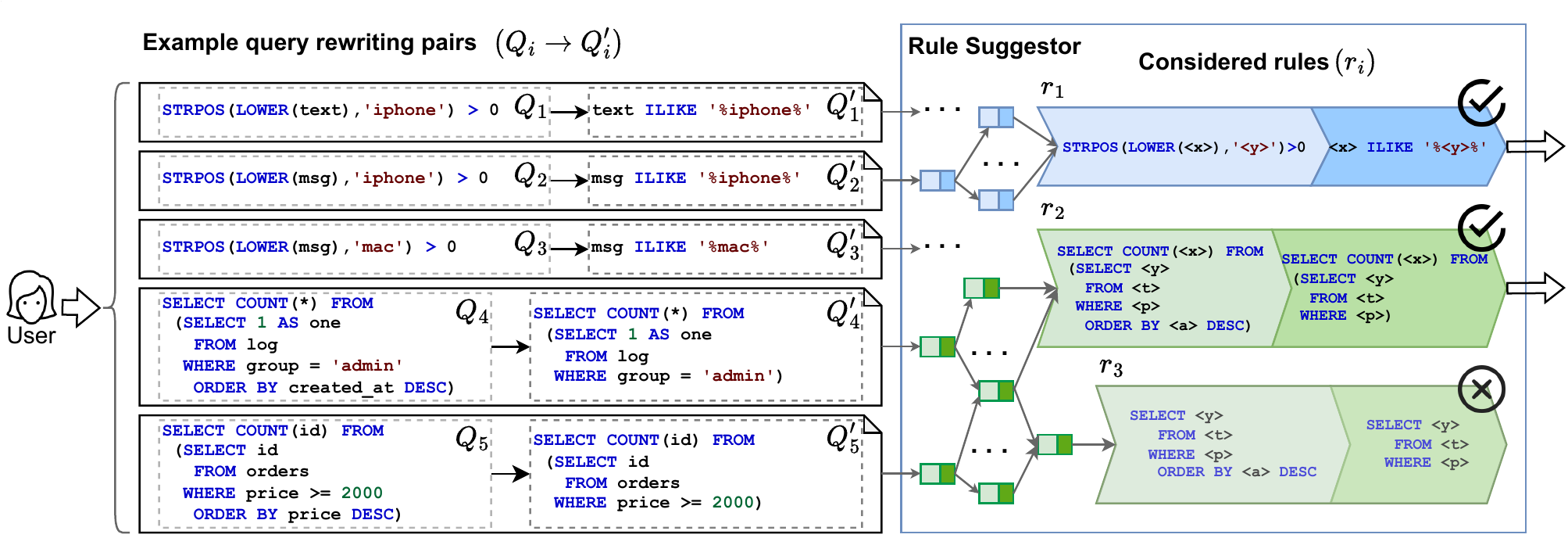}
  \caption{Suggesting rewriting rules from user-given examples.  The rule suggestor suggests two rewriting rules ($r_1$ and $r_2$) that cover all five query rewriting pairs provided by the user, and the total description length of $r_1$ and $r_2$ is minimized compared to other suggestions.}
  \label{fig:rule-suggestor}
  \vspace{-2pt}
\end{figure*}

%\boldstart{Adopting MDL principle to measure the quality of rules.}  
When the rule suggestor generates rules from the user-given examples, there can be many different sets of rules that can achieve the example rewritings.  For instance, consider the five input examples in Figure~\ref{fig:rule-suggestor}.  The rule suggestor can output the original five rewriting pairs as five rules to the user.  Apparently, this suggestion is an overfit to the given examples since the suggested rules cannot rewrite queries slightly different from the examples.  Intuitively, we want to suggest more general rules that capture the pattern of the given examples.  At the same time, we do not want to over-generalize the rules, which may underfit the examples.  For instance, in Figure~\ref{fig:rule-suggestor}, both rules $r_2$ and $r_3$ can achieve the rewritings for the example pairs $(Q_4, Q_4')$ and $(Q_5, Q_5')$, which removes the \texttt{ORDER  BY} clause from the subquery.  In this case, $r_2$ is less general than $r_3$ but is a better suggestion, because lacking the context of a \texttt{COUNT} aggregation in the outer query, $r_3$ can be erroneous in many cases.  

To this end, we want to avoid underfitting or overfitting the given examples when measuring the quality of rewriting rules.  An effective way is through the Minimum Description Length (MDL) principle~\cite{journals/automatica/Rissanen78}, which minimizes the total length required to describe the underlying patterns in the data.  There are  MDL-based metrics for pattern extractions in domains such as data mining~\cite{journals/datamine/GarofalakisGRSS03}, data cleaning~\cite{conf/vldb/RamanH01, journals/pvldb/HeCGZNC18}, and regex learning~\cite{conf/cikm/BrauerRMB11}.  We can adapt these existing metrics to measure our rewriting rules' quality or derive our own description length functions as needed.  From the rule-suggestor's perspective, we assume a rule-quality metric is given.  

\rtwo{
For the MDL metric, we assume no access to the target database.  If we are granted access, we can also consider the rewriting rules' effectiveness in improving the performance of the historical workload as the rules' quality.  For simplicity, we first use MDL as the quality function and then discuss how to extend the framework to include query performance to measure the rules' quality in Section~\ref{sec:adding-query-cost-to-rule-quality}.
}

\boldstart{Rewriting-rule suggestion problem.}  Next, we formally define the problem of suggesting high-quality rules from given examples.

\begin{definition}{(Covering) \label{def:covering}}
  Let $Q$ be a set of query rewriting pairs $\{(Q_1,Q_1')$ $,(Q_2,Q_2')$ $,\ldots$ $,(Q_n,Q_n')\}$, and $R$ be a set of rewriting rules $\{r_1$ $,r_2$ $,\ldots$ $,r_k\}$.  We say $R$ {\em covers} $Q$ if for each pair $(Q_i,Q_i')$ in $Q$, there is at least one rule $r_j$ in $R$ such that $r_j$ can rewrite $Q_i$ into $Q_i'$, and there is no rule $r_k$ in $R$ such that $r_k$ can rewrite $Q_i$ into a query different than $Q_i'$. 
\end{definition}

\begin{definition}{(Rewriting-rule suggestion problem)}
\label{def:problem-formulation}
  Let $Q$ be a given set of query rewriting pairs $\{(Q_1,Q_1')$ $,(Q_2,Q_2')$ $,\ldots$ $,(Q_n,Q_n')\}$, $G$ be a given rule language, and $L$ be a given description length function.  The {\em rewriting-rule suggestion problem} is to compute a set $R$ of rewriting rules $\{r_1$ $,r_2$ $,\ldots$ $,r_k\}$ written in $G$ such that $R$ covers $Q$ and the total length of rules $\Sigma_{i=1\ldots k}L(r_i)$ is minimal.
\end{definition}

We propose a two-step solution.  First, we define a set of transformations that can generalize a rewriting rule into a more general form such that the transformed rule can cover more rewriting pairs than the original rule.  By applying the transformations on the given rewriting pairs iteratively, we identify a set of candidate rules to consider for the final suggestion.  Second, we adopt a greedy-search strategy to efficiently explore different subsets of rules as candidates and minimize the total description length.  Next, we present the technical details of both steps.

\subsection{Transforming Rules to More General Forms}
\label{sec:rule-transformation}

\begin{figure}[htb]
  \centering
  \includegraphics[width=.99\linewidth]{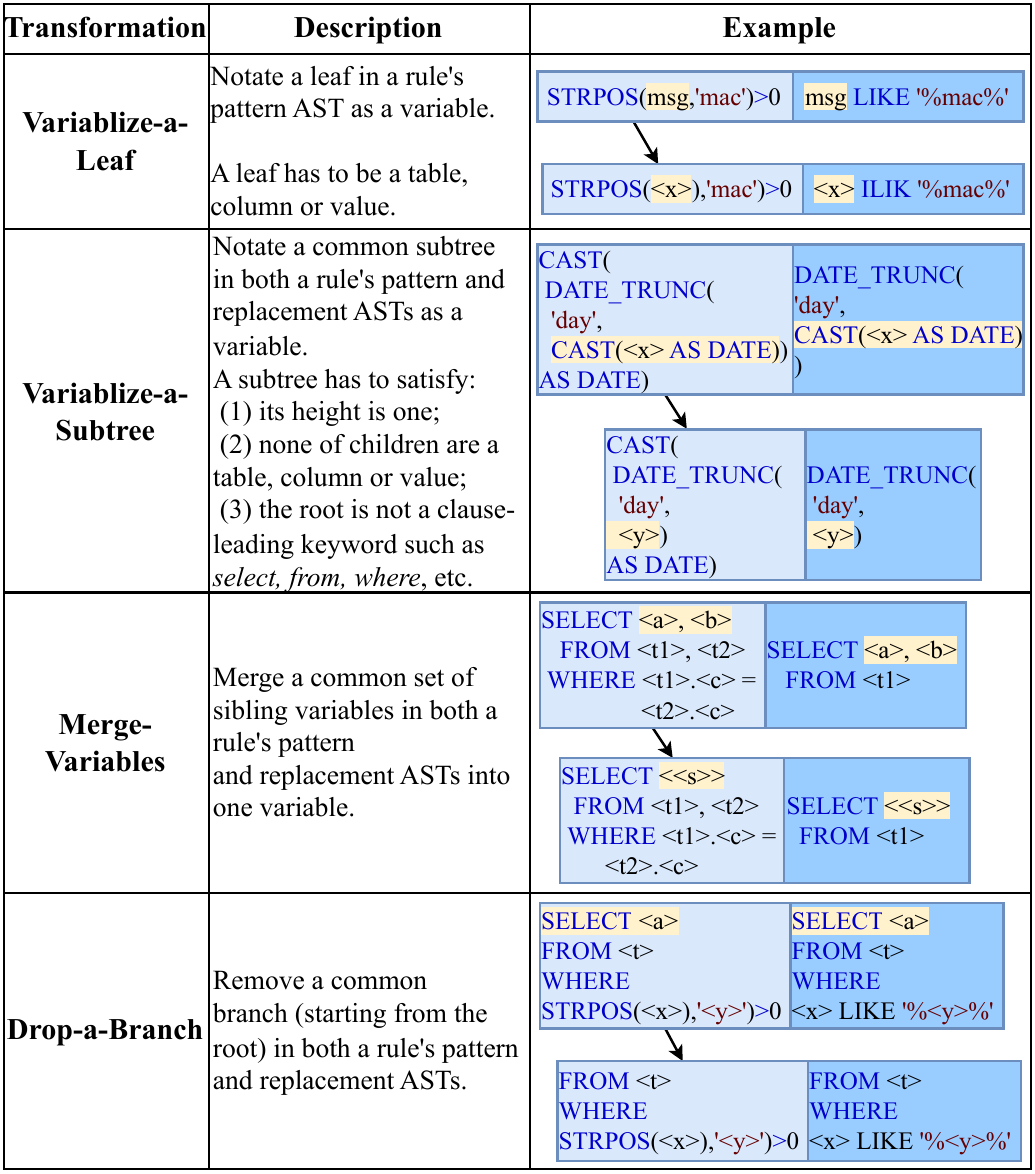}
  \caption{Transformations on rewriting rules formulated in \langname.  A transformation is applied to the pattern and replacement ASTs of a rewriting rule to generalize it into a more general rule. }
  \label{fig:transformations}
  \vspace{-6pt}
\end{figure}

\boldstart{Transformations on rules.}  A transformation on a rewriting rule can generalize the rule into a more general rule such that the new rule covers more rewriting pairs than the original one.  The instantiation of transformations is dependent on the given rule language.  We now define transformations (shown in Figure~\ref{fig:transformations}) on rewriting rules formulated in the \langname language, namely \textit{Variablize-a-Leaf}, \textit{Variablize-a-Subtree}, \textit{Merge-Variables}, and \textit{Drop-a-Branch}.  The last three transformations only happen if the replaced variables are not referred to in other places in the rule's pattern or replacement. 

\boldstart{Variablize-a-Leaf.}  This transformation replaces an instantiated element (table, column, or value) in a rule with a variable.  In this way, the transformed rule can match more queries than the original one.  As shown in the first example in Figure~\ref{fig:transformations}, the transformed rule can match a query with any column name in the first argument of the \texttt{STRPOS} function. In contrast, the original rule can only match a query with the specific ``msg'' column.

\boldstart{Variablize-a-Subtree.}  This transformation replaces a complex element (expression, predicate, or subquery) in a rule with a variable.  In this way, we can generalize the pattern of a rule by hiding the details within an expression, predicate, or subquery.  In the second example in Figure~\ref{fig:transformations}, the common expression ``\texttt{CAST(<x> AS DATE)}'' appears without any modifications in both the rule's pattern and replacement, which means that it might be an irrelevant pattern in the original rule.  Summarizing the common expression with a new variable makes the rule more general.

\begin{figure}[htb]
  \centering
  \includegraphics[width=.95\linewidth]{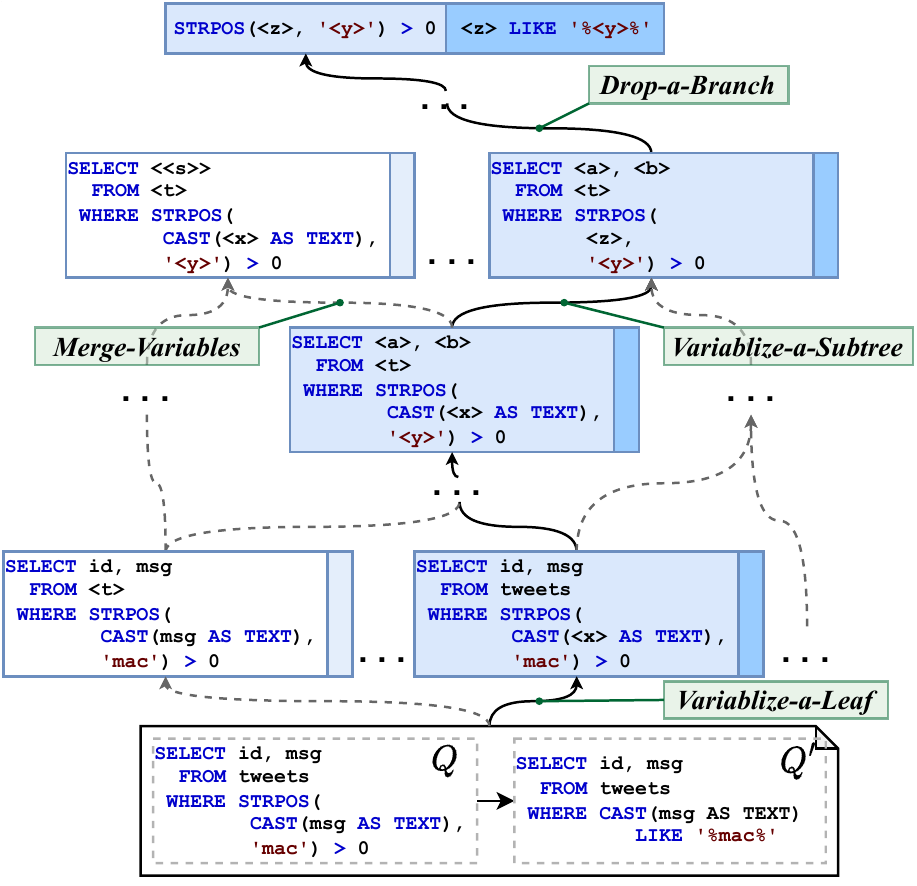}
  \caption{A rule graph generated from a given rewriting pair ($Q$, $Q'$).  The vertices are generalized rules (only showing patterns due to the space limit).  The solid edges show one path of generalizing the pair into a general rule.  The green tags on the edges illustrate which transformations are applied.}
  \label{fig:rule-graph}
\end{figure}

\boldstart{Merge-Variables.}  Notice that in the \langname language, an element-variable can only match a single element in queries.  We introduce this transformation to generalize a set of variables to a set-variable to suppress the quantity restriction when matching queries.  As shown in the third example in Figure~\ref{fig:transformations}, the original rule only matches queries with the two columns in the \texttt{SELECT} clause, and the transformed rule can match queries with any number of columns in the selection list.  This transformation is useful when we want a more general rule where the quantity of elements does not matter for the pattern.

\boldstart{Drop-a-Branch.}  This transformation is a complement of the \textit{Variablize-a-Subtree} transformation.  Since \langname requires the pattern of a rule to be a valid full or partial SQL query, we cannot variablize an entire clause.  For example, in the fourth example in Figure~\ref{fig:transformations}, if we variablize the \texttt{SELECT <a>} subtree as a new variable \texttt{<y>}, the transformed pattern ``\texttt{<y> FROM <t> WHERE \ldots}'' is not valid SQL syntax.  Thus, we introduce the \textit{Drop-a-Branch} transformation, which removes a common branch in a rule's pattern and replacement.  In this way, we can gradually remove the irrelevant context of a rule's pattern from the top to the bottom of the pattern's AST.

\boldstart{Rule Graph.}  Until now, starting from an initial rewriting pair, by applying the transformations iteratively, we can generalize it into more and more general rewriting rules gradually.  If we treat each newly-generated rewriting rule as a vertex and a transformation as an edge, we can obtain a graph of rewriting rules.  We call it a {\em rule graph}.  Figure~\ref{fig:rule-graph} shows an example rule graph.  As we can see, a rule graph for a single rewriting pair can be big, and the union of all rule graphs for a set of rewriting pairs can be even bigger, so searching for a set of high-quality rules is difficult.  In the next section, we discuss how to navigate through the search space and make final suggestions to the users.

\section{Searching For High-Quality Rules \label{sec:searching-high-quality-rules}}

To solve the rewriting-rule suggestion problem defined in Definition~\ref{def:problem-formulation}, we defined a set of transformations in Section~\ref{sec:rule-transformation} to generalize the initial rewriting pairs to more general rewriting rules.  However, the candidate sets of generalized rules that can cover the initial rewriting examples may be large.  It can be computationally expensive to search all possible sets to compute an optimal solution with the minimum description length.  To solve the problem, we adopt a heuristic-based strategy to expand the candidate-rule set greedily and rely on a local set of rules to make final suggestions.  In this section, we first present the greedy searching framework, then propose several heuristics to further reduce the search overhead.

\subsection{A Greedy Searching Framework}

We develop a method to search for rules, as shown in Algorithm~\ref{alg:greedy}.  Its main idea is the following.  We start with the original rewriting pairs as a basic solution, and treat each query pair as a rewriting rule (line~\ref{alg:greedy:init}).  We iteratively replace rules in the solution with a more general rule that reduces the total description length the most.  In each iteration, we first explore a set of candidate rules by applying transformations to the rules in the current solution (line~\ref{alg:greedy:explore}).  We say a rule $x$ {\em covers} another rule $y$ if $x$'s pattern matches $y$'s pattern and $x$ can rewrite $y$'s pattern to $y$'s replacement.  For each candidate rule, we compute the reduction of the total description length if we use it to replace its covered rules in the solution (lines~\ref{alg:greedy:for-begin}-\ref{alg:greedy:for-end}).  
We then choose the rule that has the maximum reduction (line~\ref{alg:greedy:choose}) and replace its covered rules with the new rule (line~\ref{alg:greedy:replace}).  We stop the iteration if there is no more reduction (line~\ref{alg:greedy:stop}).  In this case, we return the current solution (line~\ref{alg:greedy:return}).

\begin{algorithm}[htb]
\caption{A greedy algorithm for suggesting rules \label{alg:greedy}}
    \SetKwRepeat{Do}{do}{while}
    \DontPrintSemicolon
    \KwIn{A set of rewriting pairs $\mathcal{Q}=\{(Q_1, Q_1'), \ldots, {(Q_n, Q_n')}\}$ \newline 
          A set of transformations $\mathcal{T}=\{T_1, T_2, \ldots, T_m\}$ \newline
          A description length function $\mathcal{L}$ on a rule
    }
    \KwOut{A set of rewriting rules $\mathcal{R}$}
    
    $\mathcal{R}$ $\leftarrow$ $\mathcal{Q}$\;\label{alg:greedy:init}
    \While{True}{
        $\mathcal{C}$ $\leftarrow$ \textbf{\texttt{Explore\_Candidates}}$(\mathcal{R}, \mathcal{T})$\;\label{alg:greedy:explore}
        \For{$c$ $\in$ $\mathcal{C}$}{ \label{alg:greedy:for-begin}
            \tcp*[h]{find rules that can be replaced by $c$}\;
            $\mathcal{R}_c$ $\leftarrow$ $\{R_i$ $\in$ $\mathcal{R}$ $|$ $R_i$ is covered by $c\}$\;
            \tcp*[h]{compute the length reduction if $c$ replaces $\mathcal{R}_c$}\;
            $\Delta\mathcal{L}_c$ $\leftarrow$ $\sum_{R_i\in\mathcal{R}_c}\mathcal{L}(R_i)$ $-$ $\mathcal{L}(c)$\; \label{alg:greedy:benefit}
        }\label{alg:greedy:for-end}
        \tcp*[h]{choose a candidate rule with the largest length reduction}\;
        $\hat{c}$ $\leftarrow$ $\argmax_{c \in \mathcal{C}}{\Delta\mathcal{L}_c}$\;\label{alg:greedy:choose}
        \tcp*[h]{stop when there is no more reduction}\;
        \If{$\Delta\mathcal{L}_{\hat{c}}$ $\leq$ $0$}{\label{alg:greedy:stop}
            \Return $\mathcal{R}$\;\label{alg:greedy:return}
        }
        \tcp*[h]{update the result set}\;
        $\mathcal{R}$ $\leftarrow$ $\mathcal{R}$ $-$ $\mathcal{R}_{\hat{c}}$ $+$ $\hat{c}$\;\label{alg:greedy:replace}
    }
\end{algorithm}

The algorithm follows the hill-climbing paradigm~\cite{books/aw/RN2020}, where in each iteration, it explores a set of candidate rules to consider as the possible next directions.  The exploration of candidates is implemented in the {\em Explore\_Candidates($\mathcal{R}$,$\mathcal{T}$)} procedure, and the decision of which set of candidates to explore can affect how easily the algorithm is stuck at a local optimum.  Ideally, the explored candidates should include all possible rules transformed from the current rule set.  However, the size of the transformed rules can be large. Thus, we need to consider the trade-off between the exploration size and the probability of trapping in a local optimum.  We discuss different methods in the following. 

\boldstart{A naive candidate-exploration method.}  A naive method is to parameterize the number of hops when we transform the rules in the given rule set.  As shown in the rule graph in Figure~\ref{fig:rule-graph}, starting from a base rule, we can transform it into different child rules by applying different transformations.  We call a child rule a ``$1$-hop rule'' if it is obtained from the base rule by applying one transformation.  Similarly, a rule is a ``$k$-hop rule'' if it is obtained after applying $k$ transformations on the base rule one by one.  The parameter $k$ decides the exploration overhead of the searching framework.  We can increase $k$ to allow the algorithm to look ahead before settling down at a local optimum at a higher computational cost.  We call this method ``$k$-hop-neighbor exploration'' (KHN for short).

This method has two problems.  One is that it is hard to decide the $k$ value.  A $k$ value may be good for some input examples but can be bad for others.  The second problem is that a fixed $k$ value for all base rules ignores their different amounts of potential to discover a high-quality rule.  To solve these two problems, we propose an adaptive exploration method next.

\subsection{Exploring Candidate Rules Adaptively}

In this subsection, we discuss how to explore candidates in an adaptive way by considering the different amounts of potential of transforming different base rules to discover a high-quality general rule.  The goal is to explore more promising candidate rules first to fill a fixed size of the candidate set.

\boldstart{$m$-promising neighbors.}  Its main idea
%, as shown in Figure~\ref{fig:adaptive-exploration}, 
is that instead of exploring neighbors a fixed number of hops away from the current rule set, we explore a fixed number (denoted as $m$) of neighbors that can reduce the total length of the rule set the most.  The value $m$ directly decides the computation overhead of the rule-suggestion algorithm.  We can decide its value by considering the running time (e.g., 2 seconds) allowed to run the algorithm and the hardware resources we have.  To find the $m$ neighbors, we explore the given base rule set iteratively. In each iteration, we choose a rule that is most promising to be transformed into a more general rule that reduces the total length the most.  In this way, we can generate a set of candidate rules with different numbers of hops transformed from different base rules in the given rule set.  

%% First figure backed up in overflow.tex

Algorithm~\ref{alg:adaptive-exploration} shows the pseudo-code of the method of $m$-promising-neighbor exploration (MPN for short).  For a given function $\mathcal{P}$ that measures a rule's {\em promisingness score}, the algorithm starts from the initial rule set, chooses one rule with the highest score, replaces it with all its $1$-hop transformed rules in the candidate rule set, and stops until the rule set reaches the given size $m$. 

\begin{algorithm}[htb]
\caption{$m$-promising-neighbor exploration \label{alg:adaptive-exploration}}
    \SetKwRepeat{Do}{do}{while}
    \DontPrintSemicolon
    \KwIn{A set of rewriting rules $\mathcal{R}=\{R_1, R_2, \ldots, R_n\}$ \newline 
          A set of transformations $\mathcal{T}=\{T_1, T_2, \ldots, T_m\}$ \newline
          A function $\mathcal{P}$ that measures a rule's promisingness score \newline
          A parameter $m$ that limits the output size
    }
    \KwOut{A set of candidate rewriting rules $\mathcal{C}$}
    
    $\mathcal{C}$ $\leftarrow$ $\mathcal{R}$\;
    \While{$|\mathcal{C}|$$<$$m$}{
        \tcp*[h]{choose the most promising candidate rule}\;
        $\hat{c}$ $\leftarrow$ $\argmax_{c \in \mathcal{C}}{\mathcal{P}(c)}$\;
        \tcp*[h]{replace it with its $1$-hop transformed child rules}\;
        \For{$T_i$ $\in$ $\mathcal{T}$}{ 
            $T_i(\hat{c})$ $\leftarrow$ $\{$all possible child rules by applying $T_i$ to $\hat{c}\}$\;
            $\mathcal{C}$ $\leftarrow$ $\mathcal{C}$ $\cup$ $T_i(\hat{c})$\;
        }
        $\mathcal{C}$ $\leftarrow$ $\mathcal{C}$ $-$ $\hat{c}$\;
    }
    \Return $\mathcal{C}$\;
\end{algorithm}

\boldstart{Measuring the promisingness score of a rule.}  We consider three signals to measure a rule's promisingness score.  First, one signal is the total length of those base rules that can be covered if we transform a candidate rule into a more general form.  Second, another signal is the number of transformations needed to apply to a candidate rule if we want it to cover more base rules in the rule set.  This signal measures how far we can reach a more general rule starting from a particular candidate rule.  Third, the last signal is the length of a candidate rule.

Formally, suppose we are given a set of base rewriting rules $\mathcal{R}=\{R_1, R_2, \ldots, R_n\}$ and a candidate rule $c$.  We compute rule $c$'s promisingness score $\mathcal{P}(c)$ as follows.  For each base rule $R_i$ $\in$ $\mathcal{R}$, we compute a distance $\mathcal{D}(c, R_i)$, which is the number of transformations on rule $c$ to cover rule $R_i$.  We will discuss how to compute this value shortly.  Let $\mathcal{L}$ be the given description length function.  The promisingness score of rule $c$ is:
$$
\mathcal{P}(c)=\sum_{i=1}^{n} \frac{\mathcal{L}(R_i)}{\mathcal{D}(c, R_i)} + \frac{1}{\mathcal{L}(c)}.
$$

If a rule can be generalized with fewer transformations to cover longer base rules and its own length is shorter, it should have a higher promisingness score.  We now describe how to compute the distance $\mathcal{D}(c, R_i)$ of transforming rule $c$ to cover the base rule $R_i$.  We count the number of transformations on $c$ to produce a more general form $c'$ to cover rule $R_i$.  A rule $c'$ covers rule $R_i$ if the pattern of $c'$ matches $R_i$'s pattern, and we can rewrite it to $R_i$'s replacement.  Therefore, we can run the pattern-matching process of $c$ on $R_i$ similar to that of a rule on a query.  The only difference is that when we find any mismatching part, instead of immediately returning false, we compute the number of transformations needed for the mismatching part in $c$ to match that in $R_i$. 

%% Last paragraph and last figure backed up in overflow.tex

\subsection{\rtwo{Including Query Cost in Rule Quality} \label{sec:adding-query-cost-to-rule-quality}}

\rtwo{
To this end, we use MDL as a metric for rewriting rules' quality.  We now show how to include the effectiveness in improving the performance of a historical workload $\mathcal{W}$ to measure the rules' quality.  For a given candidate rewriting rule set $\mathcal{R}$ (in Algorithm~\ref{alg:greedy}), we can obtain a set $\mathcal{W}_{\mathcal{R}}$ of rewritten queries by rewriting $\mathcal{W}$ using the rules in $\mathcal{R}$.  Suppose we know the cost of queries to the target database. We can obtain the total cost of all rewritten queries in $\mathcal{W}_{\mathcal{R}}$, denoted as $\mathcal{C}(\mathcal{W}_{\mathcal{R}})$.  When we evaluate the benefit of replacing a few rules in $\mathcal{R}_c$ with a candidate rule $c$ (line~\ref{alg:greedy:benefit}), we compute the reduction of query cost when using the new rule set to rewrite $\mathcal{W}$, denoted as $\Delta \mathcal{C}_c$.  Then, we compute a weighted sum of both the reduction of description length and the reduction of query cost as the total benefit for the candidate rule $c$ as 
$$
Benefit_{c} \leftarrow \beta \times \frac{\Delta \mathcal{L}_c}{\mathcal{L}_{\mathcal{R}}} + (1 - \beta) \times \frac{\Delta \mathcal{C}_c}{\mathcal{C}(\mathcal{W}_{\mathcal{R}})},
$$
where $\beta$ is a parameter to tune the balance between the importance of the description length and performance improvement of the rewriting rules.  We replace the original $\Delta \mathcal{L}_c$ with the new $Benefit_{c}$ at lines~\ref{alg:greedy:benefit}, \ref{alg:greedy:choose}, and \ref{alg:greedy:stop}, and extend Algorithm~\ref{alg:greedy} to include the effectiveness in improving workload performance to measure the quality of rewriting rules.  Similarly, we also include the new benefit value when computing the promisingness score of a rule in Algorithm~\ref{alg:adaptive-exploration}.}

\section{Experiments \label{sec:experiments}}

We conducted experiments to evaluate \sysname regarding three aspects: formulating rules using the \langname rule language, suggesting rules from user-given examples, and the end-to-end performance using \sysname to rewrite queries.  In particular, we want to answer the following questions: (1) How easy is it for SQL developers to use the \langname language to formulate query rewriting rules?  (2) What is the expressive power of  \langname? (3) Are the transformations defined in \sysname enough to generate general rules from example pairs?  (4) How do different search strategies perform in terms of running time and rule qualities?  (5) How much benefit can \sysname provide on the end-to-end query performance with the human-centered query rewritings?

\subsection{Setup}

\boldstart{Workloads.}  We used four workloads as shown in Table~\ref{tab:workloads}. Each workload had a set of SQL rewriting pairs, and each pair consisted of an original query and a rewritten query.  Each rewritten query was equivalent to and usually outperformed its original query.  The WeTune workload included $245$ pairs of SQL queries published in the appendix table in the paper~\cite{conf/sigmod/WangZYDHDT0022}.  They collected those original queries from $20$ open source applications on GitHub and generated the rewritten queries by applying their machine-discovered rewriting rules.  The Calcite workload comprised $232$ rewriting pairs of SQL queries designed for the Apache Calcite test suite~\cite{website:calcite_tests}.  

To consider the real-world use cases where business intelligence (BI) users do interactive analysis on their data residing in a database, we created \rmeta{three more workloads using Tableau~\cite{tableau:tableau} and Apache Superset~\cite{Superset}} on top of PostgreSQL and MySQL.  The ``Tableau + TPC-H'' workload included $20$ rewriting pairs of SQL queries, which corresponded to the top $20$ queries in the TPC-H benchmark~\cite{misc/TPC-H}.  We first inserted a $10$GB TPC-H synthesized dataset into a PostgreSQL database (indexes were created using Dexter~\cite{website:dexter}),  then used Tableau Desktop software to connect to the PostgreSQL database in its live mode.  For each query in the TPC-H benchmark, we manually built a Tableau visualization workbook that could answer the corresponding business question, then collected the backend SQL query generated from Tableau for the workbook.  We then analyzed the Tableau-formulated SQL query and came up with a rewritten query with a better performance.  Similarly, we generated the ``Tableau + Twitter'' \rmeta{and ``Superset + Twitter'' workloads} by building visualization dashboards \rmeta{using Tableau and Apache Superset} to analyze 30 million tweets on their textual, temporal, and geospatial dimensions on top of both Postgres and MySQL databases.  In the workloads, \rmeta{$14$} pairs of queries were generated on top of PostgreSQL, and \rmeta{$11$} pairs were generated on top of MySQL.

\begin{table}[htbp]
\small
\caption{Workloads used in the experiments.}
\label{tab:workloads}
\begin{tabular}{|l|l|r|l|}
\hline
\bf{Id} & \bf{Workload} & \bf{\# of query pairs} \\ \hline
1 & Calcite & 232 \\ \hline
2 & WeTune & 245  \\ \hline
3 & Tableau+TPC-H & 20 \\ \hline
4 & Tableau+Twitter & \rmeta{14} (Postgres) + \rmeta{6} (MySQL) \\ \hline
\rmeta{5} & \rmeta{Superset+Twitter} & \rmeta{5 (MySQL)} \\ \hline
\end{tabular}
\end{table}

\boldstart{Testbed.}  We implemented the \sysname system  using Python $3.9$ and used the ``mo-sql-parsing'' package~\cite{website:mo-sql-parsing} as the SQL parser.  All experiments were run on a MacBook Pro $2017$ model with a $2.3$GHz Intel Core i$5$ CPU, $8$GB DDR$3$ RAM, and $256$GB SSD.  The Tableau Desktop software version was $2021.4$\rmeta{, and the Apache Superset version was May $2023$}.  The PostgreSQL software version was $14$, and the MySQL software version was $8.0$. 

\boldstart{Description length function.}  To evaluate the performance of the rule-suggestion algorithms, we implemented a description length function designed for rules rewritten in \langname.  We followed the design principles proposed in~\cite{conf/vldb/RamanH01}.  The main idea was that each rule had a constant basic length, and the more variables it had, the larger its description length should be.  In this case, the description length metric made sure that high-quality rules could match as many given examples as possible, but they were not over-generalized to match unseen queries.  We computed the description length $\mathcal{L}$ of a particular rule $r$ as the following.  Let $W$ be the constant basic length of any rule, $W_E$ be the weight of an element-variable, and $W_S$ be the weight of a set-variable.  We used three counters in the given rule. We counted the number of element-variables in the rule as $C_E$ and the number of set-variables as $C_S$. In addition, we counted the number of non-variable elements in the rule as $C_O$, where non-variable elements included keywords, values, table names, column names, etc. In the end, we computed the length $\mathcal{L}$ of rule $r$ as 
$$
\mathcal{L}(r) = W + (W_E \times C_E + W_S \times C_S) / C_O.
$$

\subsection{A User Study to Evaluate Rule Languages}

\begin{table}[htbp]
\small
\caption{User profiles in the user study.}
\label{tab:user-study-profiles}
\begin{tabular}{|r|r|r|r|r|r|}
\hline
\multicolumn{1}{|l|}{{\bf{Background}}} & Faculty & Staff & \shortstack[l]{Software \\ Engineers} & \shortstack[l]{Ph.D.\\ students} & \shortstack[l]{M.S.\\ students} \\ \hline
\multicolumn{1}{|l|}{{\bf{\% of users}}} & 4.5\% & 4.5\% & 4.5\% & 72.7\% & 13.6\% \\ \hline
\end{tabular}
\end{table}

We conducted a user study to evaluate how easy it was for SQL users to use  \langname to formulate rewriting rules.  Besides \langname, we considered two other languages for comparison.  One was regular expression~\cite{wiki:regular-expression}, and we used its C Sharp implementation provided by regex101~\cite{website:regex101}.  The other was the internal rule language used by WeTune~\cite{conf/sigmod/WangZYDHDT0022}, and we used its own implementation provided by its demo website WeRewriter~\cite{website:WeRewriter}.  We selected three rewriting pairs of SQL queries from two workloads on two databases.  One pair was from the WeTune workload, and the other two pairs were from the ``Tableau + Twitter'' workload on both PostgreSQL and MySQL.  For each rewriting pair, we showed the original and rewritten queries to the user, along with three rewriting rules defined in the three languages that could achieve the same rewriting.  We asked the user to ``\emph{select one of the three rules that you think is the easiest to understand.}''  In the questions, we randomized the orders of the rule languages and hid their names to make the comparison fair.

\begin{table}[htbp]
\small
\caption{Results in the user study (\% of users selected the rule language as the easiest to understand).}
\label{tab:user-study-results}
\begin{tabular}{|r|r|r|r|}
\hline
\multicolumn{1}{|l|}{{\bf{Pair Id}}} & 1 & 2 & 3 \\ \hline
\multicolumn{1}{|l|}{{\bf{Workload}}} & Twitter(Postgres) & Twitter(MySQL) & WeTune(Q91) \\ \hline
\multicolumn{1}{|l|}{{\bf{\% of Regex}}} & 13.6\% & 4.5\% & 0\% \\ \hline
\multicolumn{1}{|l|}{{\bf{\% of WeTune}}} & 0\% & 13.6\% & 13.6\% \\ \hline
\multicolumn{1}{|l|}{{\bf{\% of \langname}}} & \bf{86.4\%} & \bf{81.8\%} & \bf{86.4\%} \\ \hline
\end{tabular}
\end{table}

We invited $22$ users who were familiar with SQL and with different backgrounds.  The profiles of users are shown in Table~\ref{tab:user-study-profiles}, and the results are summarized in Table~\ref{tab:user-study-results}.  Among all the rewriting pairs, more than $80\%$ of users selected the rule formulated in \langname as the easiest to understand, and it outperformed the other two languages significantly.  The user study results show that \langname is an easy-to-use language and was preferred by SQL users.

\subsection{Comparison of Rule-Searching Strategies}
We evaluated the performance of different searching strategies in the rule-suggestion searching framework.  We compared the three strategies discussed in Section~\ref{sec:searching-high-quality-rules}.  The first was ``Brute-Force'' (``BF'' for short), which explored all possible rules that were transformed from the current rule set in the {\em Explore\_Candidates} procedure.  The second was the ``$k$-hop-neighbor exploration'' (``KHN'' for short), where we explored the neighbors of a fixed number ($k$) of hops away from the base rules for each iteration's consideration.  The last was the adaptive exploration method, ``$m$-promising-neighbor exploration'' (``MPN'' for short), where we explored a fixed number ($m$) of neighbors that were the most promising to finally reduce the total description length of the resulting rule set.  We used the ``Tableau + Twitter'' workload and varied the number of rewriting examples as the input to the searching algorithms.  For each input set of examples, we first ran the BF method to get a high-quality set of suggested rules as the benchmark.  We then ran the KHN and MPN methods and made sure they both output the same set of suggested rules as the BF method by gradually increasing the $k$ and $m$ parameters.  In this way, we ensured the fairness of the comparison between different methods.

\begin{figure}[htbp]
  \centering
  \begin{subfigure}[t]{0.495\linewidth}
    \includegraphics[width=\linewidth]{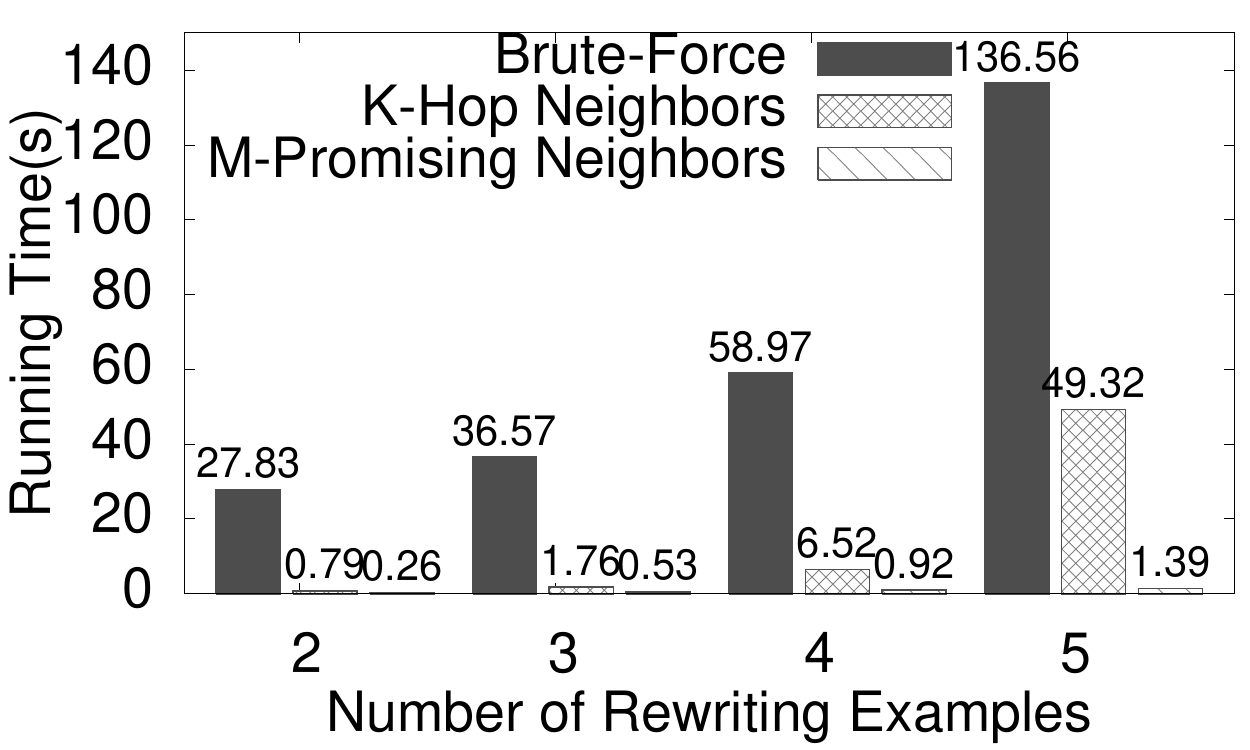}
    \caption{Running time. \label{fig:rule-suggestion-run-time}}
  \end{subfigure}
  \hfill
  \begin{subfigure}[t]{0.495\linewidth}
    \includegraphics[width=\linewidth]{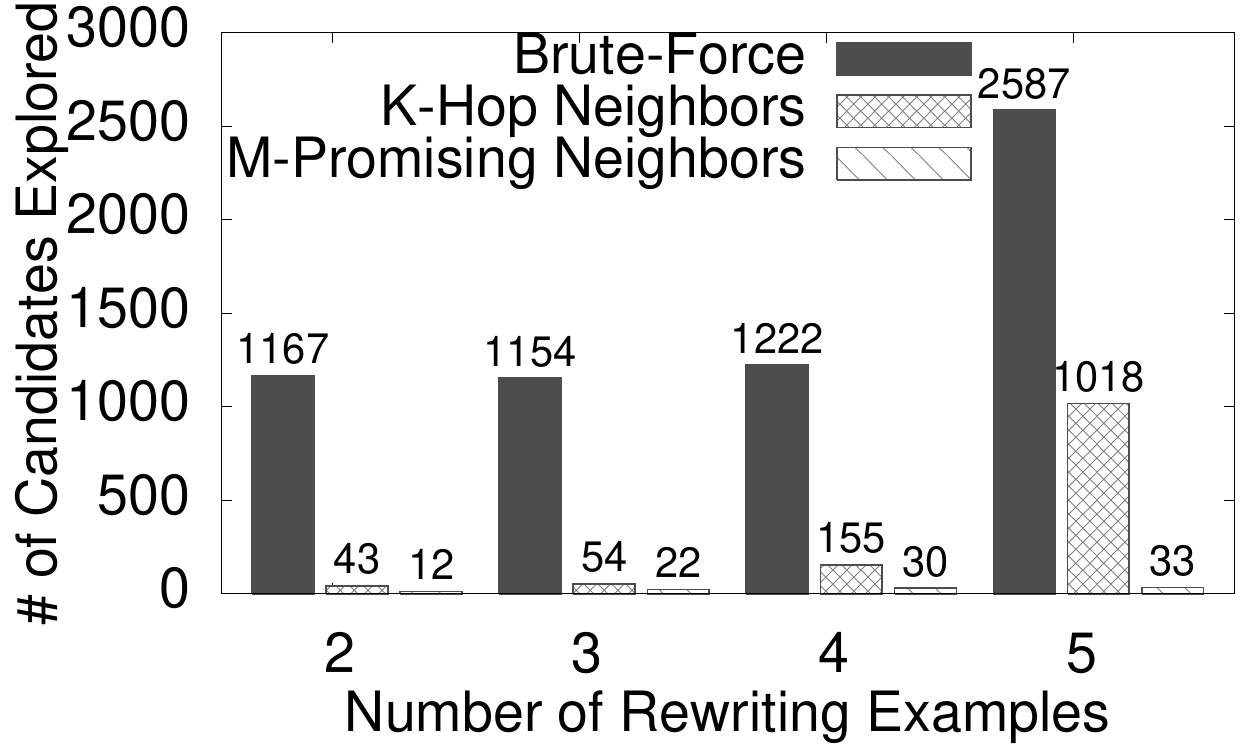}
    \caption{Total \# of candidates explored. \label{fig:rule-suggestion-cnt-candidates}}
  \end{subfigure}
  \caption{Comparison of different candidate exploration methods to suggest the same set of rules on the ``Tableau + Twitter'' workload. \label{fig:rule-suggestion-comparison}}
  \vspace{-6pt}
\end{figure}

The results are shown in Figure~\ref{fig:rule-suggestion-comparison}.  As shown in Figure~\ref{fig:rule-suggestion-run-time}, as the number of input examples increased, the running time of the brute-force method increased sub-exponentially.  The reason was for each example added to the input set, the number of candidate rules generated from the new example was exponential to its number of elements in the original query.  Compared to the brute-force method, the KHN method had significantly less running time since it only explored a small set of candidate rules during the exploration phase.  However, its running time still went up to $50$ seconds for $5$ input examples.  The reason was that to reach the high-quality rules, the KHN method had to tune its $k$ value to $4$, and the number of explored rules increased exponentially with the increase of the $k$ value.  In comparison, the MPN method outperformed both other methods significantly, and the running time increased linearly as the input set size increased.  These results are consistent with those shown in Figure~\ref{fig:rule-suggestion-cnt-candidates}, and both figures illustrate the correlation between the running time and the number of candidates explored in the searching framework.

\subsection{Effect of $m$ in $m$-promising Neighbors}

\begin{figure}[htb]
    \centering
    \begin{subfigure}[t]{0.495\linewidth}
        \includegraphics[width=\linewidth]{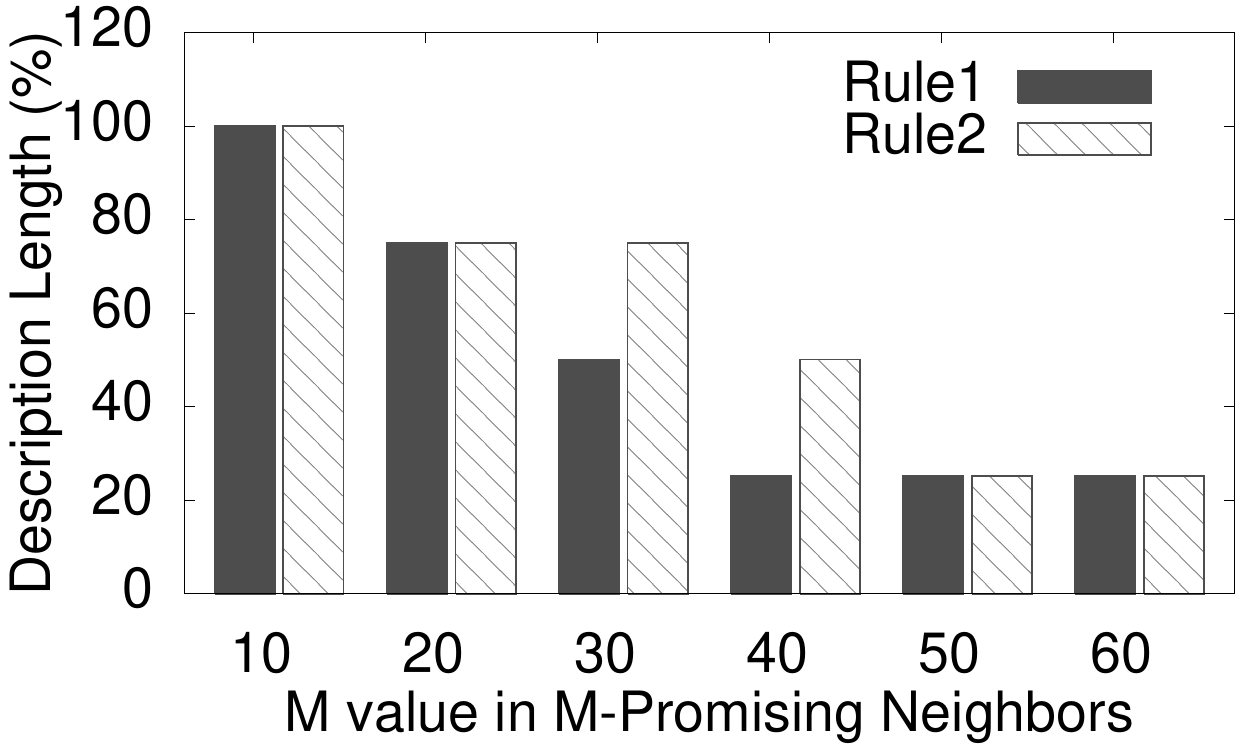}
        \caption{Description Length (\% of the raw examples). \label{fig:mpn-description-length}}
    \end{subfigure}
    \hfill
    \begin{subfigure}[t]{0.495\linewidth}
        \includegraphics[width=\linewidth]{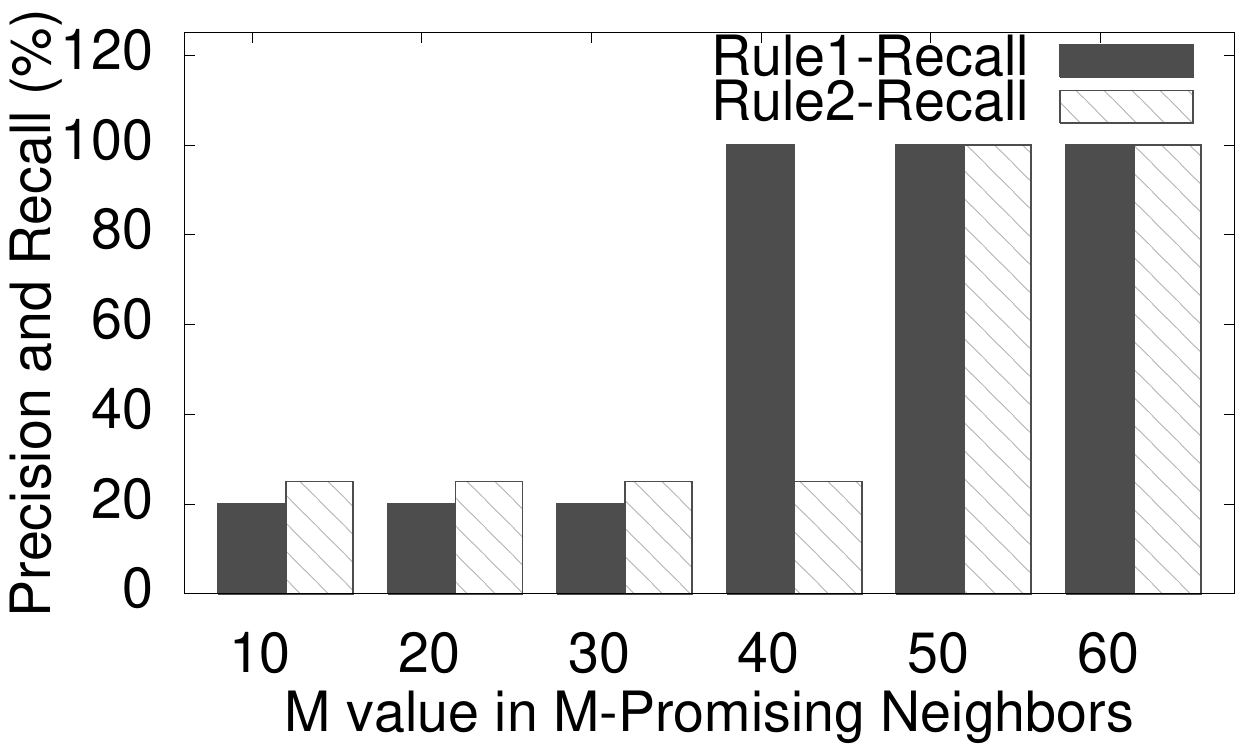}
        \caption{Precision and Recall on unseen pairs. \label{fig:mpn-precision-recall}}
    \end{subfigure}
    
    \begin{subfigure}[t]{0.495\linewidth}
        \includegraphics[width=\linewidth]{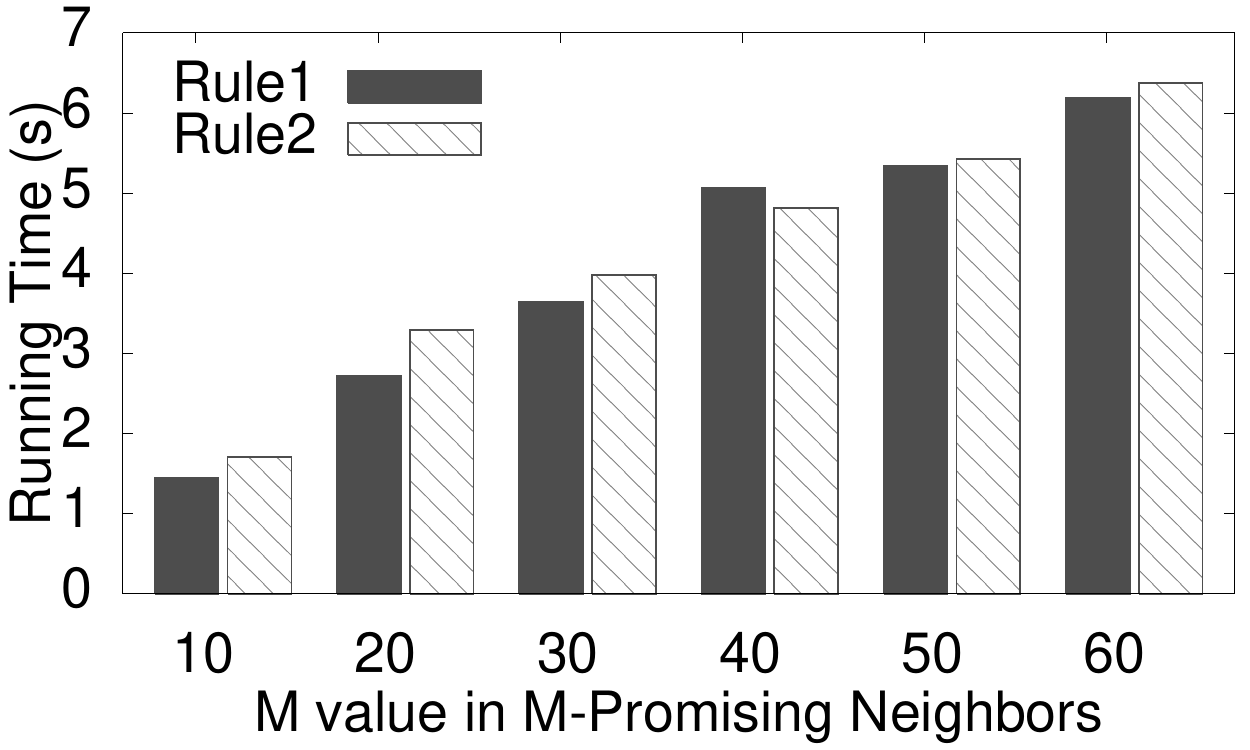}
        \caption{Running Time (s). \label{fig:mpn-run-time}}
    \end{subfigure}
    \hfill
    \begin{subfigure}[t]{0.495\linewidth}
        \includegraphics[width=\linewidth]{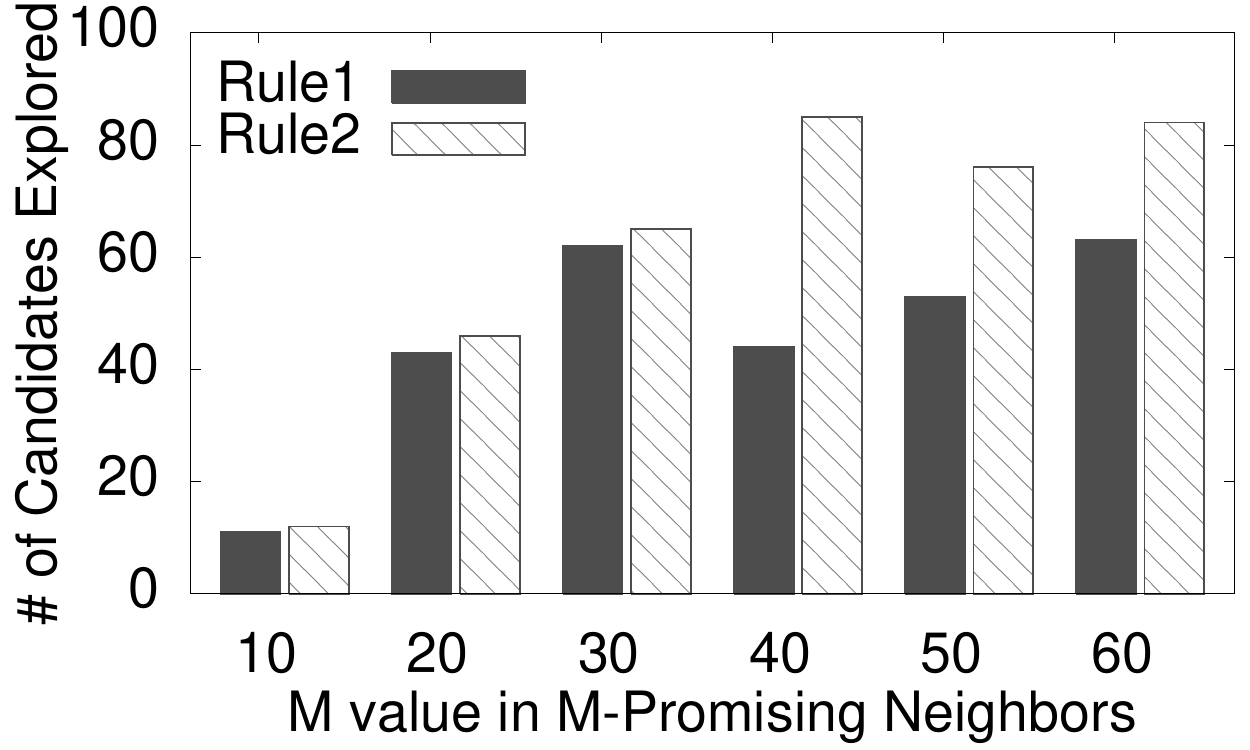}
        \caption{Total \# of candidates explored. \label{fig:mpn-cnt-candidates}}
    \end{subfigure}
    \caption{Effect of the $m$ value in the $m$-promising-neighbor searching strategy on the WeTune workload. \label{fig:mpn-m-effect}}
    \vspace{-6pt}
\end{figure}

We evaluated the effect of the $m$ value in the $m$-promising-neighbor searching strategy on the WeTune workload.  We randomly chose $30$ rewriting pairs within the first three applications in the workload as the testing set.  We then chose the top two frequent rewriting patterns and named them as ``{\em Rule1}'' and ``{\em Rule2}''.  Among the $30$ pairs, there were $5$ pairs matching {\em Rule1} and $4$ pairs matching {\em Rule2}.  For each rule, we used one matching pair as the seed and manually generated $4$ rewriting examples as the input examples for the rule-suggestion algorithm.  We ran the algorithm using the $m$-promising-neighbor strategy with different $m$ values.  We measured the total description length of the output rule set and the result is shown in Figure~\ref{fig:mpn-description-length}.  It shows that for both rules' input example sets when the $m$ value increased, the output of the rule-suggestion algorithm converged to the optimal rule set with the minimum description length.  Referring to the corresponding running time shown in Figure~\ref{fig:mpn-run-time}, it only took about $5$ to $6$ seconds for the algorithm to output the optimal rule set.  Figure~\ref{fig:mpn-cnt-candidates} also shows the numbers of candidates explored for different $m$ values.

We evaluated the output rule set from the rule-suggestion algorithm on the unseen $30$ testing rewriting pairs in the workload.  We measured both the precision and recall computed as follows.  Suppose the rule set rewrote $x$ unseen pairs of queries, among which $x1$ pairs satisfied the intent of the user. Then the precision is $\frac{x1}{x}$.  Suppose the user wanted $y$ pairs of queries in the testing set to be successfully rewritten, and the rule set only rewrote $y1$ out of $y$. Then the recall is $\frac{y1}{y}$.  The result is shown in Figure~\ref{fig:mpn-precision-recall}.  The precision was always $100\%$ (omitted in the figure) because the design of the description length function enforced that the rules were never over-generalized.  And the recall was initially low for a small $m$ value because the output rules were very specific to the input examples, and the output rules were not optimal yet.  As the $m$ value increased to $50$ or more, the algorithm started to output the optimal suggested rules that could cover unseen query pairs with similar patterns, which led to a $100\%$ recall in the end.

\subsection{End-to-End Query Time Using \sysname \label{sec:end-to-end-query-time}}

We evaluated the end-to-end query time (the time between the frontend sending the SQL query to and receiving the result from the database) using \sysname to rewrite queries in the \rmeta{``Tableau + TPC-H'' and ``Tableau + Twitter'' workloads on PostgreSQL and the ``Superset + Twitter'' workload on MySQL}.  For each query in the workload, besides the running time of the original query formulated by \rmeta{Tableau or Superset on PostgreSQL or MySQL}, we also collected the running time of two rewritten queries using different rewriting rules.  One rewritten query (noted as ``{\em Rewritten Query (WeTune Rules)}'') was obtained from the WeRewriter~\cite{website:WeRewriter} system, which used rewriting rules automatically discovered by WeTune.  The other rewritten query (noted as ``{\em Rewritten Query (Human Rules)}'') was obtained from \sysname using human-crafted rewriting rules based on manual analysis of the original query and its physical plan.

\begin{figure}[htb]
  \centering
  \includegraphics[width=0.99\linewidth]{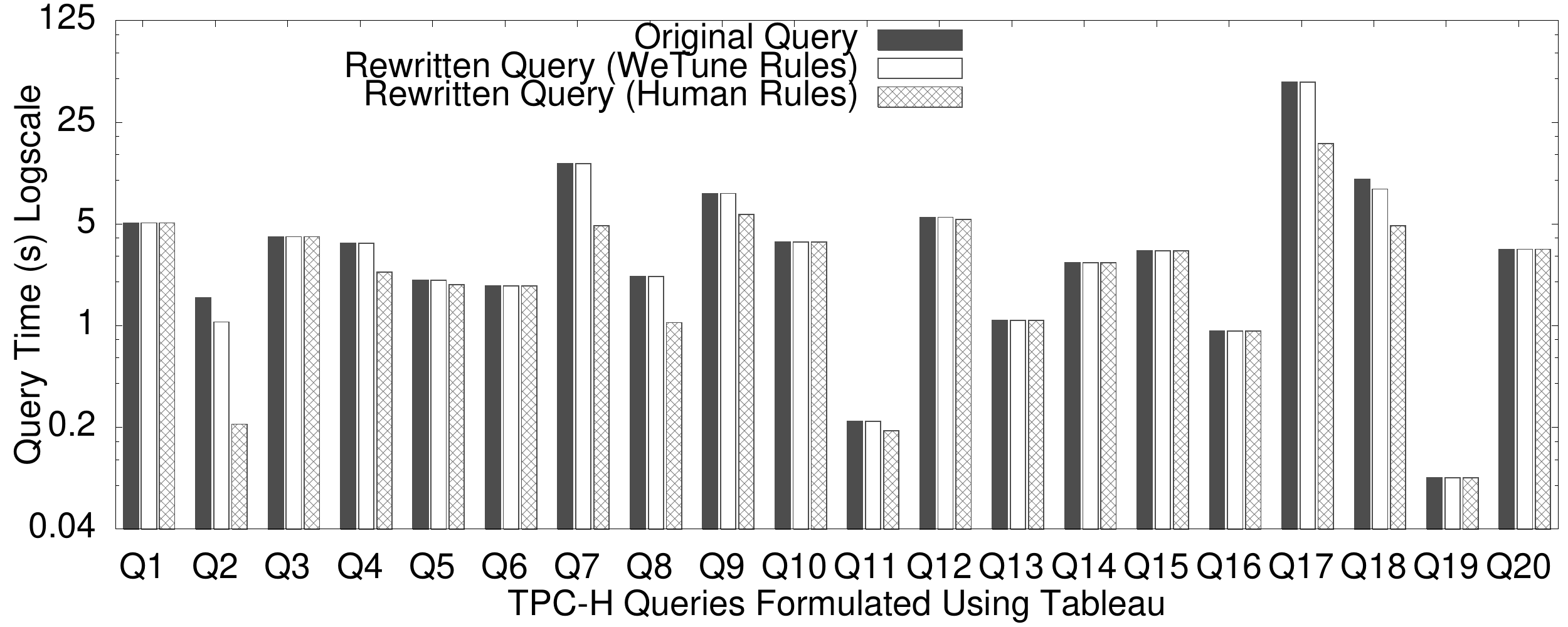}
  \caption{End-to-end query time using \sysname to rewrite queries with WeTune-generated rules and human-crafted rules on ``Tableau + TPC-H'' workload compared to original query time in PostgreSQL.}
  \label{fig:tpch-query-time}
  \vspace{-6pt}
\end{figure}

\rmeta{Figure~\ref{fig:tpch-query-time} shows the result for the workload of ``Tableau + TPC-H'' on Postgres.}  Among the $20$ queries, only two rewritten queries ($Q2$ and $Q18$) using the WeTune-generated rules could reduce the query time.  At the same time, using human-crafted rewriting rules, \sysname reduced $10$ queries' running time, which comprised $50\%$ of all the queries.  Within the $10$ rewritten queries using human-crafted rules, $70\%$ of them reduced the original queries' running time significantly (by more than $25\%$).  For example, $Q2$ was reduced by $86\%$ ($1.555$s to $0.207$s) and $Q17$ was reduced by $61\%$ ($47.046$s to $17.802$s).  \rone{Note that in the $10$ queries optimized using human-crafted rules, $7$ of them used statement-level reshaping rules such as ``join-to-exists'', ``remove-subquery'', etc., and $3$ of them used hints such as ``force-join-order''.} %This result illustrated the importance of human-centered query rewriting.

\begin{figure}[htbp]
  \centering
  \begin{subfigure}[t]{0.495\linewidth}
    \includegraphics[width=\linewidth]{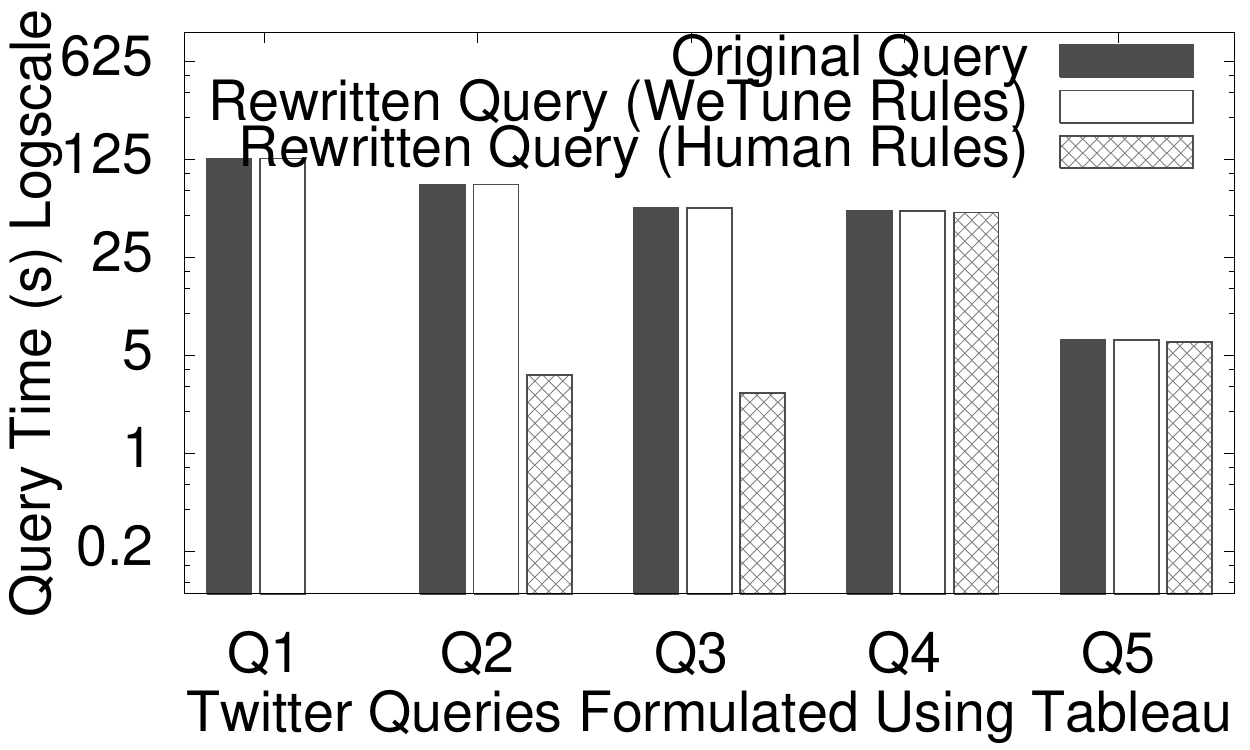}
    \caption{\rmeta{``Tableau + Twitter''.} \label{fig:tableau-twitter-query-time}}
  \end{subfigure}
  \hfill
  \begin{subfigure}[t]{0.495\linewidth}
    \includegraphics[width=\linewidth]{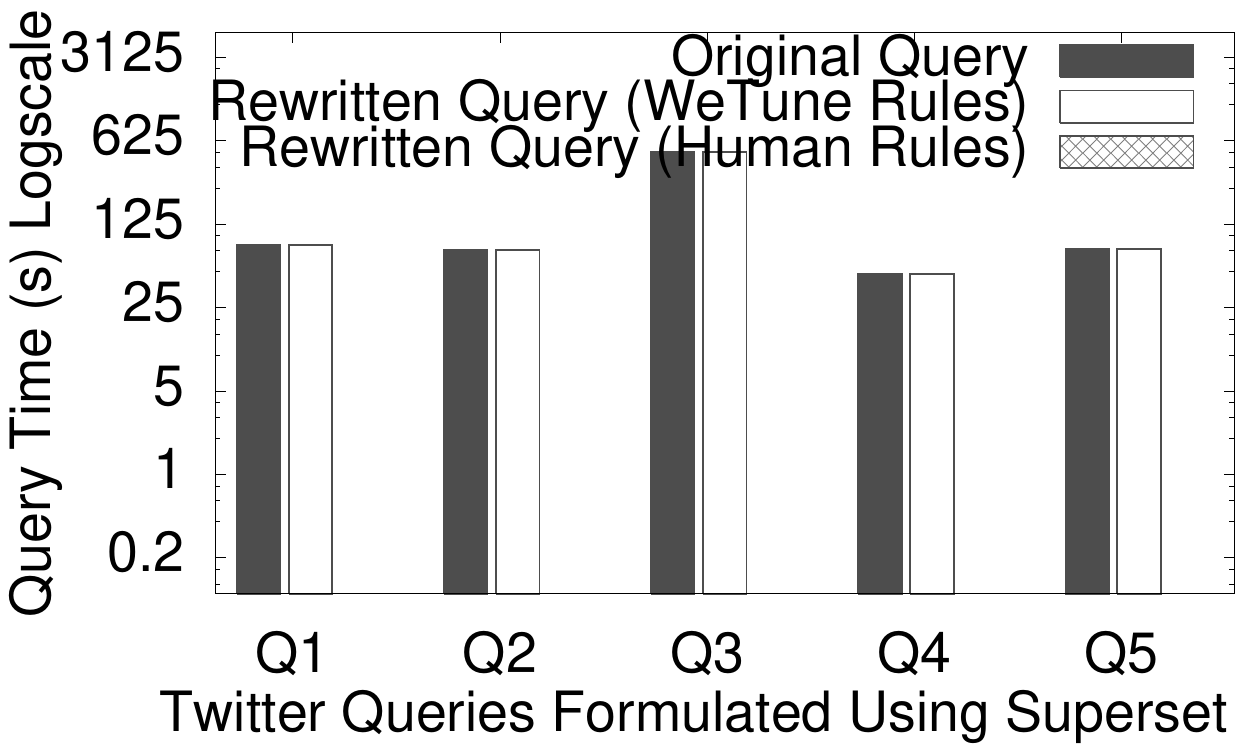}
    \caption{\rmeta{``Superset + Twitter''.} \label{fig:superset-twitter-query-time}}
  \end{subfigure}
  \caption{\rmeta{End-to-end query time using \sysname to rewrite queries with WeTune-generated rules and human-crafted rules compared to original query time in MySQL.} \label{fig:twitter-mysql-query-time}}
  \vspace{-6pt}
\end{figure}

\rmeta{Figure~\ref{fig:tableau-twitter-query-time} and~\ref{fig:superset-twitter-query-time} show the results for the workloads of ``Tableau + Twitter'' and ``Superset + Twitter'' on MySQL.  For the $5$ Tableau queries on MySQL, the human-crafted rewriting rules were mainly predicate-level removing unnecessary {\tt ADDDATE} calculation, as discussed in Section~\ref{sec:insufficiency-of-existing-solutions}.}  \rtwo{For the $5$ Superset queries on MySQL, the human-crafted rewriting rules were mainly translating the textual filtering condition from a {\tt LIKE} predicate to a full-text search predicate since MySQL does not support any index-scan for {\tt LIKE} predicate but does for full-text search. For example, in one query, the human-crafted rule rewrote the predicate ``{\tt text} {\tt LIKE} {\tt '\%stopasian} {\tt \-hate\%'}'' to ``{\tt MATCH(text)} {\tt AGAINST} {\tt ('stopasianhate} {\tt stopasian} {\tt \-hatecrime} {\tt stopasianhatecrimes')}, which was equivalent only for this particular dataset because all substring ``stopasianhate'' matched records could be matched using the three full-text keywords: ``stopasianhate'', ``stopasianhatecrime'', and ``stopasianhatecrimes''.}  \rmeta{As shown in Figure~\ref{fig:superset-twitter-query-time}, all $5$ queries were accelerated by $100+$ times (e.g., $83$s to $0.8$s) due to this human-crafted rewriting rule, which shows the importance of the proposed human-centered query rewriting approach.}

\subsection{Generality of Rule Transformations}

To evaluate the generality (covering more rewriting examples) of rule transformations, we used $178$ rewriting pairs in the Calcite workload after removing some examples that the third-party SQL parser failed to parse.  We applied the transformations defined in Section~\ref{sec:rule-transformation} to each rewriting example iteratively as long as they are applicable to generate more general rules.  If multiple examples were generalized to the same rule, we only kept one copy of the rule.  We divided the transformations into $5$ categories: ``\textit{variablize-a-table}'', ``\textit{variablize-a-column}'', ``\textit{variablize-a-value}'', ``\textit{variablize-a-subtree}'', and ``\textit{merge-variables}'', and we gradually increased the number of categories applied in the rule generalization process.  We compared the rewriting result of the generated rules on the $178$ examples.  Since most rewriting pairs in the Calcite workload were designed for a unique rule, most of the examples were rewritten by the rules generated from themselves.  We collected the percentage of examples that were rewritten by rules generated by other examples and named it ``{\em sharing-rule examples (\%)}''.  We also collected the precision and recall of using the generalized rules to rewrite the original queries of the input examples.

\begin{figure}[htbp]
  \centering
  \begin{subfigure}[t]{0.495\linewidth}
    \includegraphics[width=\linewidth]{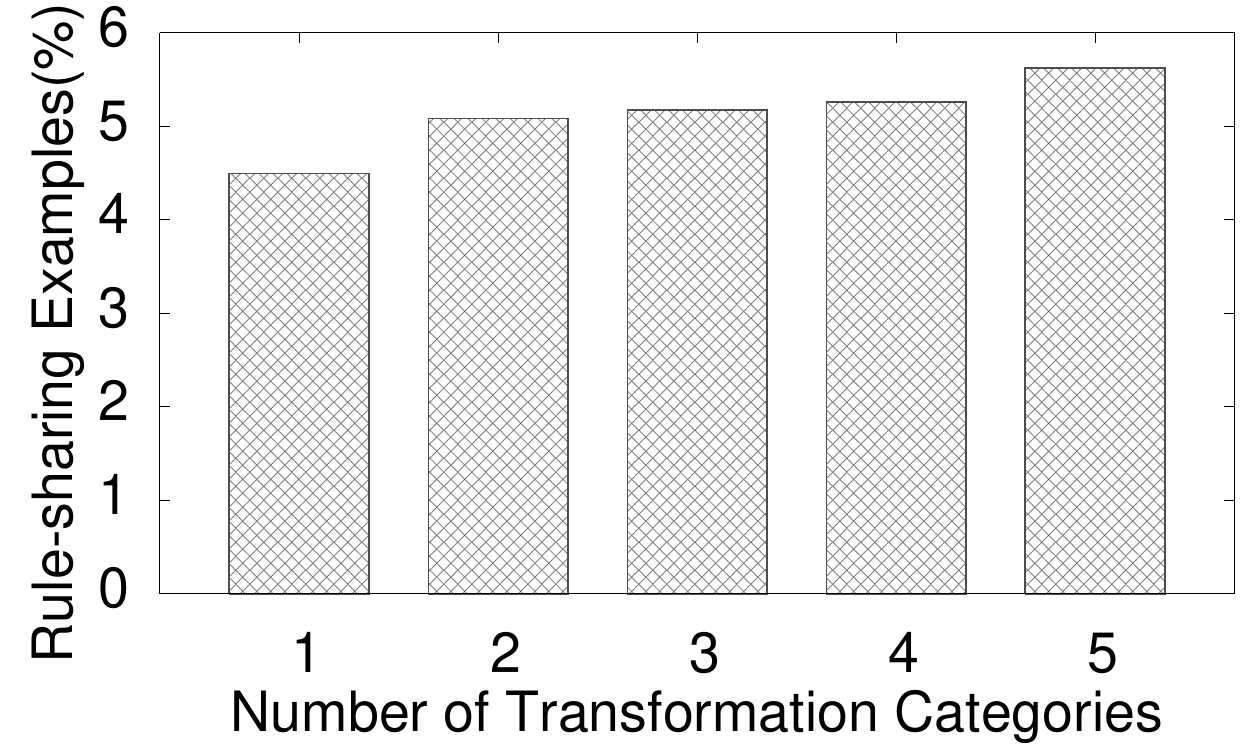}
    \caption{Rule sharing percentage. \label{fig:calcite-rule-sharing-percentage}}
  \end{subfigure}
  \hfill
  \begin{subfigure}[t]{0.495\linewidth}
    \includegraphics[width=\linewidth]{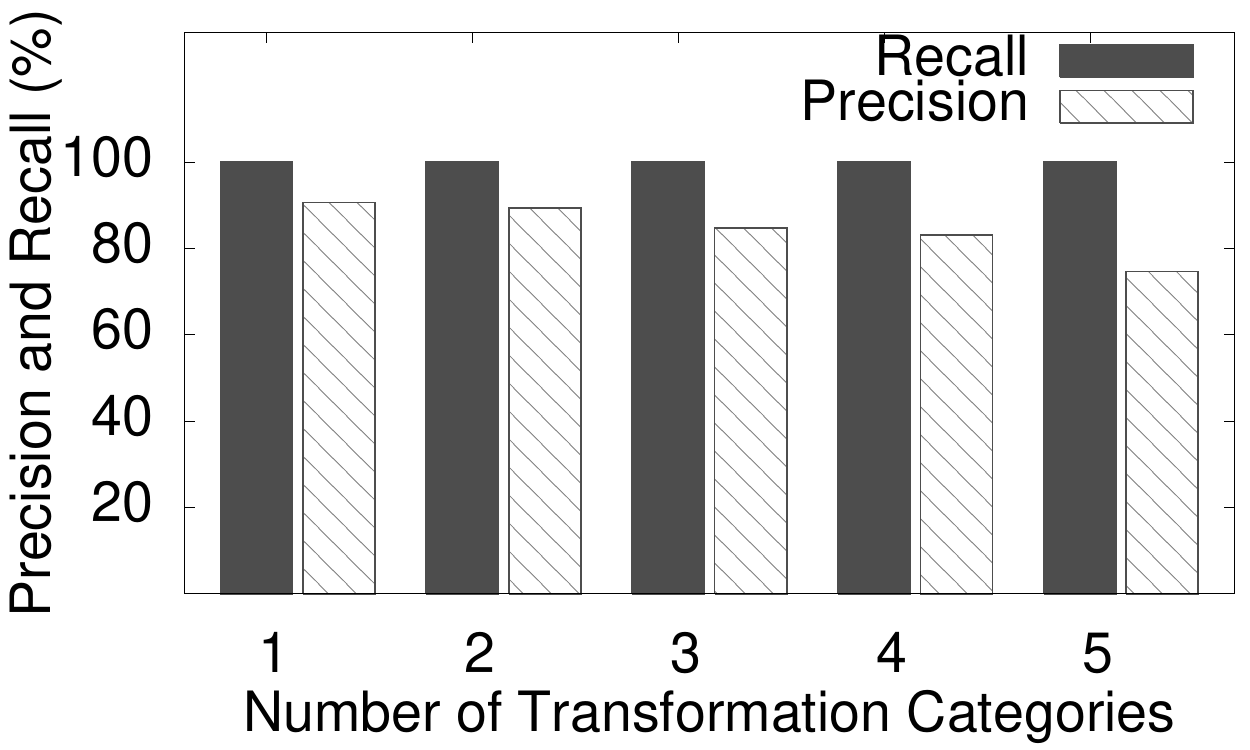}
    \caption{Precision and Recall. \label{fig:calcite-precision-recall}}
  \end{subfigure}
  \caption{The generality of rules generalized from the Calcite examples using different sets of transformations. \label{fig:calcite-generality}}
  \vspace{-6pt}
\end{figure}

Figure~\ref{fig:calcite-rule-sharing-percentage} shows that with more transformation categories used in generating rules, there were more examples rewritten by rules generated from other examples, which means the generated rules were more general.  Figure~\ref{fig:calcite-precision-recall} shows that with more transformations used to generalize rules, the recall remained $100\%$ because more general rules could always match the seed examples.  However, the rewriting precision went down. The reason was that the more transformations resulted in over-generalized rules that matched examples they should not match and rewrote the original queries to unintentional forms.  This result also motivated our consideration of using MDL as an objective function to suggest rules, which prevented over-generalization for given examples.

\subsection{\rtwo{Effect of Different Rule Quality Metrics}}

\rtwo{
We also evaluated the effect of using different importance weights $\beta$ in the {\em benefit} value (defined in Section~\ref{sec:adding-query-cost-to-rule-quality}) using the ``Tableau + Twitter'' workload.  We used four query pairs as input examples for the rule-suggesting framework and another five queries as a historical workload to compute the benefit value in Algorithm~\ref{alg:greedy}.  We varied $\beta$ from $1.0$ (i.e., only considering the MDL as rule quality) to $0.0$ (i.e., only considering query cost as rule quality).  For each $\beta$ value, we first ran the rule-suggesting framework with the ``MPN'' strategy to obtain the suggested rules.  We then evaluated the suggested rules based on three metrics.  We used the rules to rewrite the five queries in the workload and collected the query ``cost reduction'' by comparing the rewritten queries' cost and the original queries' cost.  We treated the suggested rules with  $\beta=1.0$ as the baseline and then computed the ``description length increase'' and rule-suggesting algorithm ``running time increase`` for the rules suggested by other $\beta$ values.  
}

\setlength{\columnsep}{5pt}%
\setlength{\intextsep}{0pt}%
\begin{wrapfigure}{r}{0.5\linewidth}
 \centering
    \includegraphics[scale=.33]{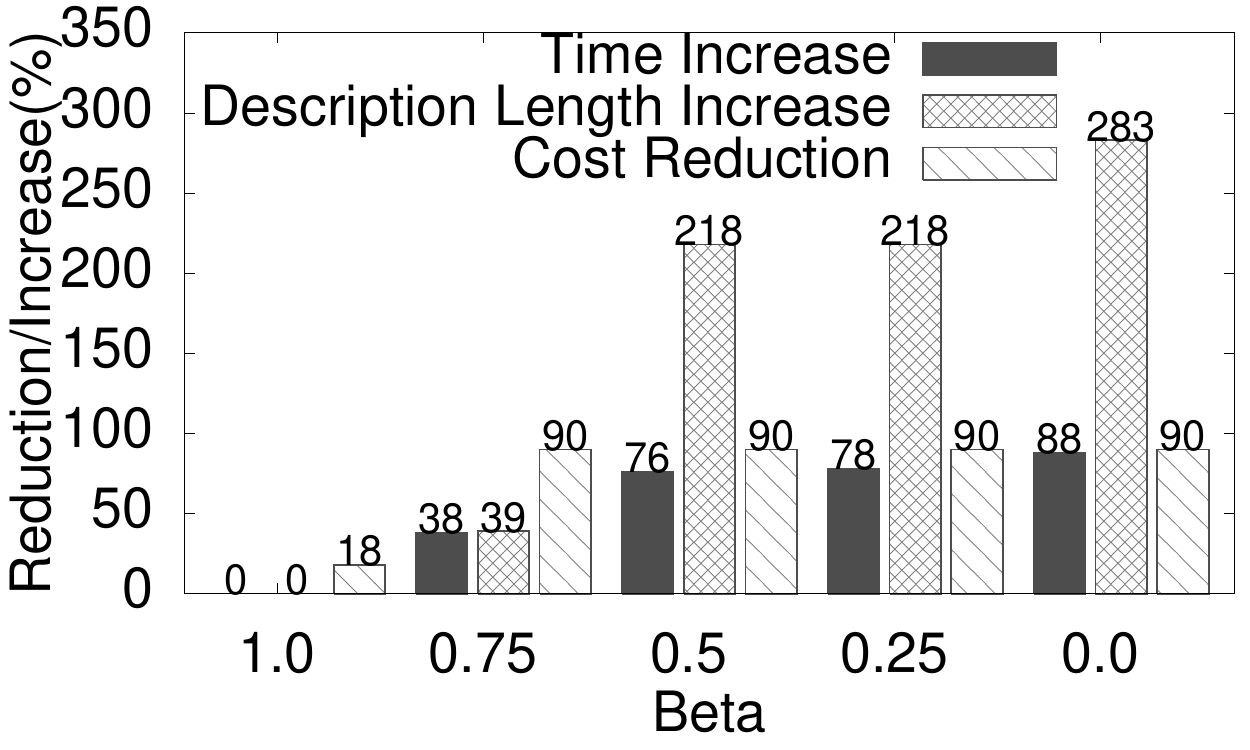}
    \caption{Effect of different $\beta$ values.}
  \label{fig:different-beta-values}
\end{wrapfigure}

\rtwo{
The results are shown in Figure~\ref{fig:different-beta-values}.  When $\beta$ was $1.0$, the suggested rules only reduced the query cost by $18\%$. When $\beta$ decreased to $7.5$, the suggested rules reduced the query cost by $90\%$.  However, the cost was both the description length of rules and the running time of the rule-suggesting algorithm increased by $40\%$.  The cost increased when $\beta$ further decreased.  When $\beta$ was $0.0$, which means the algorithm did not consider the description length at all, the total description length of suggested rules increased by $280\%$.
}

\boldstart{Remarks:}  The user study shows that more than $80\%$ SQL users preferred using the \langname rule language to formulate rewriting rules.  \sysname suggested high-quality (high precision and recall and low description length) rules from user-given examples quickly ($\leq 5$s) on different workloads.  Compared to existing query rewriting solutions with machine-discovered rewriting rules, using \sysname with human-crafted rewriting rules improved the performance of $50\%$ TPC-H queries by up to $86\%$.

\section{Conclusions}

In this paper, we proposed \sysname, a middleware service for human-centered query rewriting.  We developed a novel expressive rule language (\langname) for users to formulate rewriting rules easily.  We designed a rule-suggestion framework that automatically suggests high-quality rewriting rules from user-given examples.  A user study and experiments on various workloads show the benefit of using \langname to formulate rewriting rules, the effectiveness of the rule-suggestion framework, and the significant advantages of using \sysname to improve the end-to-end query performance.

%\clearpage

\bibliographystyle{ACM-Reference-Format}
\bibliography{references, localrefs}

\end{document}